\documentclass{aa}  

\usepackage{graphicx}
\usepackage[normalem]{ulem} 
\usepackage{txfonts}
\usepackage{amssymb}
\usepackage{mathrsfs}
\usepackage{enumerate}
\usepackage{subfigure}
\usepackage{multirow}
\usepackage[colorlinks=true,allcolors=blue]{hyperref}%
\renewcommand*\vec[1]{\ensuremath{\mathbf{#1}}}

\newcommand{\eq}[1]{Eq.~(\ref{eq:#1})}

\newcommand{\eqs}[2]{Eqs.~(\ref{eq:#1},\ \ref{eq:#2})}
 
\newcommand{\eqss}[2]{Eqs.~(\ref{eq:#1} - \ref{eq:#2})}
 \newcommand{\sect}[1]{Sect.~\ref{sec:#1}}

\newcommand{\tab}[1]{Table~\ref{tab:#1}}

\newcommand{\fig}[1]{Fig.~\ref{fig:#1}}
\newcommand{\Fig}[1]{Figure~\ref{fig:#1}}

\newcommand{\cf}{\textit{cf.} }
\newcommand{\eg}{\textit{e.g.} }
\newcommand{\ie}{\textit{i.e.} }

\newcommand{\BE}{\begin{equation}}
\newcommand{\EE}{\end{equation}}
\newcommand{\BA}{\begin{eqnarray}}
\newcommand{\EA}{\end{eqnarray}}

\newcommand{\deriv}[2]{\frac{{\mathrm d} #1}{{\mathrm d} #2}}

\newcommand{\curl}{ {\bf \nabla} \times}

\newcommand{\Nabla}{\vec{\nabla}}

\newcommand{\rmd}{{\rm d}}
\newcommand{\vol}{ {\mathcal V} }
\newcommand{\surf}{ {\mathcal S} }

\newcommand{\dV}{\, \rmd \vol}

\newcommand{\intv}{\int_{\vol}}

\newcommand{\curlA}{\Nabla \times \vA}
\newcommand{\curlAp}{\Nabla \times \vAp}

\newcommand{\divB}{\Nabla \cdot \vB}

\newcommand{\vA}{\vec{A}}

\newcommand{\vAp}{\vA_{\rm p}}

\newcommand{\vAj}{\vA_{\rm j}}

\newcommand{\vB}{\vec{B}}

\newcommand{\vBp}{\vB_{\rm p}}
\newcommand{\vBj}{\vB_{\rm j}}

\newcommand{\vj}{\vec{j}}
\newcommand{\vn}{\vec{n}}
\newcommand{\vv}{\vec{v}}

\newcommand{\E}{E_{\rm tot}}    
\newcommand{\Ep}{E_{\rm pot}}   
\newcommand{\Ej}{E_{\rm free}}  
\newcommand{\Emix}{E_{\rm mix}}  
\newcommand{\EdivBJ}{E_{\rm free,ns}} 
\newcommand{\EdivBp}{E_{\rm pot,ns}}  
\newcommand{\Efree}{E_{\rm free}}     
\newcommand{\Ediv}{E_{\rm div}}       

\newcommand{\Hm}{\mathscr{H}_{\rm m}}
\newcommand{\Hv}{H_{\rm V}}
\newcommand{\Hj}{H_{\rm j}}
\newcommand{\Hpj}{H_{\rm pj}}

\newcommand{\dHdt}{{\rm d}\Hv/{\rm d}t}
\newcommand{\fdHdt}{\deriv{\Hv}{t}}

\newcommand{\dHjdt}{{\rm d}\Hj/{\rm d}t}
\newcommand{\fdHjdt}{\deriv{\Hj}{t}}
\newcommand{\fdHjdtdiss}{\left.\fdHjdt\right|_{\rm Diss}}
\newcommand{\dHjdtdiss}{\dHjdt|_{\rm Diss}}
\newcommand{\fdHjdtown}{\left.\fdHjdt\right|_{\rm Own}}
\newcommand{\dHjdtown}{\dHjdt|_{\rm Own}}
\newcommand{\fdHjdttrans}{\left.\fdHjdt\right|_{\rm Trans}}
\newcommand{\dHjdttrans}{\dHjdt|_{\rm Trans}}
\newcommand{\dHpjdt}{{\rm d}\Hpj/{\rm d}t}
\newcommand{\fdHpjdt}{\deriv{\Hpj}{t}}
\newcommand{\fdHpjdtdiss}{\left.\fdHpjdt\right|_{\rm Diss}}
\newcommand{\dHpjdtdiss}{\dHpjdt|_{\rm Diss}}
\newcommand{\fdHpjdtown}{\left.\fdHpjdt\right|_{\rm Own}}
\newcommand{\dHpjdtown}{\dHpjdt|_{\rm Own}}
\newcommand{\fdHpjdttrans}{\left.\fdHpjdt\right|_{\rm Trans}}
\newcommand{\dHpjdttrans}{\dHpjdt|_{\rm Trans}}

\newcommand{\llnr}[1]{{\bf \color{magenta}{[]}} \color{black}} 

\begin{document}

 \title{Comparison of magnetic energy and helicity in coronal jet simulations}

 \author{
 E. Pariat\inst{1}\orcid{0000-0002-2900-0608}, 
 P. F. Wyper\inst{2}\orcid{0000-0002-6442-7818} \and 
 L. Linan\inst{3}\orcid{0000-0002-4014-1815}}

 \institute{Sorbonne Universit\'e, \'Ecole polytechnique, Institut Polytechnique de Paris, Universit\'e Paris Saclay, Observatoire de Paris-PSL, CNRS, Laboratoire de Physique des Plasmas (LPP), 75005 Paris, France \\
		 \and Durham University, Department of Mathematical Sciences, Stockton Road, Durham, DH1 3LE, UK \\
 \and Centre for Mathematical Plasma Astrophysics, Departement of Mathematics, KU Leuven, 3001 Leuven, Belgium}

 \date{Received September ; accepted}

 
 \abstract
 {While free/non-potential magnetic energy is a necessary element of any active phenomenon in the solar corona, its role as a marker of the trigger of eruptive process remains elusive. Meanwhile, recent analysis of numerical simulations of solar active events have shown that quantities based on relative magnetic helicity could highlight the eruptive nature of solar magnetic systems.   
}
 {Based on the unique decomposition of the magnetic field into potential and non-potential components, magnetic energy and helicity can also both be uniquely decomposed into two quantities. Using two 3D magnetohydrodynamics parametric simulations of a configuration that can produce coronal jets, we compare the dynamics of the magnetic energies and of the relative magnetic helicities.}
 {Both simulations share the same initial set-up and line-tied bottom-boundary driving profile. However they differs by the duration of the forcing. In one simulation, analysed in \citet{Wyper18b}, the system is driven sufficiently so that a point of no-return is passed, and that the system induces the generation of an helical jet. The generation of the jet is however markedly delayed after the end of the driving phase: a relatively long phase of lower-intensity reconnection takes place before the jet is eventually induced. 
 In the other reference simulation, the system is driven during a shorter time, and no jet is produced.}
 {As expected, we observe that the Jet producing simulation contains a higher value of non-potential energy and non-potential helicity compared to the non-eruptive system. Focussing on the phase between the end of the driving-phase and the jet generation, we note that magnetic energies remain relatively constant, while magnetic helicities have a noticeable evolution. During this post-driving phase, the ratio of the non-potential to total magnetic energy very slightly decreases while the helicity eruptivity index, that is the ratio of the non-potential helicity to the total relative magnetic helicity, significantly increases. The jet is generated when the system is at the highest value of this helicity eruptivity index. This proxy critically decreases during the jet generation phase. The free energy also decreases but does not present any peak when the jet is being generated.
}
 {Our study further strengthens the importance of helicities, and in particular of the helicity eruptivity index, to understand the trigger of solar eruptive events.}


 \maketitle
%

\section{Introduction} \label{sec:Introduction}

Understanding the physical processes at the origin of active solar events is a central problem of solar physics. Numerous and diverse models for eruptive events have been developed over time that aim to explain the different observational features of solar activity. Over the last few years, an interest on the relation between magnetic helicity and solar eruptivity has been renewed \citep[\eg reviews of ][]{Pevtsov14,Toriumi22} driven by the advances in the theory of helicity measurements\citep[\cf review sections of][]{Demoulin07,Demoulin09,Valori16} . 

Magnetic helicity, $\Hm$ (\cf. \eq{ClassicH}), quantifies the level of entanglement of the magnetic field lines in a closed magnetic system. It is a signed quantity, the classical definition of which was initially introduced by \citet{Elsasser56}. Magnetic helicity has the quasi-unique property of being an invariant of ideal magnetohydrodynamics (MHD) \citep{Woltjer58}. The concept has been later reviewed by \citet{BergerField84} and \citet{Finn85}, putting the focus on relative magnetic helicity, $\Hv$ (\cf. \eq{h}), a gauge invariant quantity which can be used to study non-magnetically closed systems, hence is more suitable for natural plasmas. Using numerical simulation, \citet{Pariat15b} confirmed the hypothesis introduced by \citet{Taylor74} that even in the presence of nonideal dynamics, the dissipation of relative magnetic helicity is negligible. Relative magnetic helicity can not be dissipated or created within the corona thus can only be transported or annihilated. This conservation properties has several major consequences, one of which possibly being that coronal mass ejections (CMEs) are the consequence of the evacuation of an excess of helicity \citep{Rust94,Low96}.

In the last ten years, robust methods have been developed \citep[see review of ][]{Valori16} that permits estimation of helicity in finite volumes \citep[\eg][]{Thalmann11,Valori12,Moraitis18}, helicity fluxes \citep[\eg][]{Dalmasse14,Pariat15b,Linan18,Schuck19}, and helicity per field line \citep[\eg][]{Russell15,Aly18,Yeates18,Moraitis19a}. Thanks to these developments, in recent years, magnetic helicity has constituted a renewed perspective to analyse and understand the generation of solar active events such as jets, flares and eruptions \citep[\eg][]{Knizhnik15,ZhaoL15,Priest16}. Different observed solar active regions have recently been investigated for their helicity content and dynamics. \citep{Valori13,Moraitis14,GuoY17,Polito17,Temmer17,James18,Moraitis19b,Thalmann19b,Thalmann21,Price19,Gupta21,Green22,Lumme22}.

Like magnetic energy, relative magnetic helicity can be decomposed when considering the potential and non-potential part of a magnetic field in a domain. \citet{Berger03} introduced the decomposition of the relative magnetic helicity into two gauge invariant components (\cf. \eq{hvdec}) : a non-potential helicity, $\Hj$ related to the current carrying magnetic field and a complementary volume-threading helicity, $\Hpj$. \citet{Pariat17} suggested that the ratio, $\eta_H$ (\cf \eq{etah}), of the current carrying helicity to the relative helicity could constitute an interesting proxy of when solar-like magnetic systems become eruptive. 

From 3D parametric simulations of solar coronal eruption \citep{Zuccarello15} driven by distinct line-tied boundary motions, \citet{Zuccarello18} studied the impact of the different driving flows on the helicity and energy injection. They found that the helicity ratio $\eta_H$ was clearly associated with the eruption trigger since the different eruptions occurred exactly when the ratio reached the very same threshold value. \citet{Pariat17} followed and estimated the helicity eruptivity index, $\eta_H$, in a set of seven simulations of the formation of solar active regions \citep{Leake13b,Leake14a}. The different simulations lead to either stable or eruptive configurations.  \citet{Pariat17} observed that the helicity ratio permitted to discriminate the two types of dynamics, stable or eruptive. \citet{Linan18} and \citet{Moraitis14} also analysed simulations in which the helicity eruptivity index presented a peak for systems leading to eruptive behavior.

These results motivated \citet{Linan18} to better understand the properties of $\Hj$ and $\Hpj$.  \citet{Linan18} provided the first analytical formulas of the time variation of non-potential and volume threading helicity. They found that the evolutions of the current-carrying and the volume threading helicities are partially controlled by a transfer term that reflects the exchange between these two kinds of helicity. This transfer term can even dominate the dynamics of non-potential helicity. The properties of the fluxes of helicities was further studied by \citet{Linan18}, along with the dynamics of the energies. \citet{Linan20} noted that magnetic helicities provided additional information to the trigger mechanism of the eruptive event comparatively to magnetic energies. Analysing the helicity flux of the simulation of \citet{Zuccarello15,Zuccarello18}, they also showed that the  threshold in the helicity eruptivity index could be reached by different evolution of $\Hj$ and $\Hpj$, implying that the way to reach the threshold was not so important as to reaching it. 

In observations, the analysis of the helicity eruptivity index  requires the knowledge of the magnetic field in the whole studied domain. As \citet{Linan18} demonstrated, $\Hj$ and $\Hpj$ cannot be estimated from their flux through the photosphere, unlike what is frequently done with relative magnetic helicity \citep[e.g as in ][]{Chae01,Nindos03,Pariat05,Pariat06a,Dalmasse13,Dalmasse14,Dalmasse18,Liokati22}. Estimates of $\Hj$ and $\Hpj$ must thus rely on magnetic extrapolation of the coronal field from photospheric measurements \citep[c.f. reviews][]{Wiegelmann12b,Wiegelmann14}. Such extrapolation must produce fields with a high degree of solenoidality for the helicity estimate to be trustworthy \citep{Thalmann19a,Thalmann19b,Thalmann20,Thalmann21,Thalmann22}. The helicity eruptivity index has thus been estimated prior to the onset of several active phenomena \citep{James18,Moraitis19b,Price19,Thalmann19b,Thalmann21,Gupta21,Lumme22}. These studies have consistently found that high values of the helicity eruptivity index are indeed indicating  the potential of active regions to produce eruptive events. On the contrary, very low values of the index were found prior to confined (CME-less) GOES X-class flares \citep{Thalmann19b,Gupta21}.
\citet{Lumme22} carried a data-driven model of build-up of of magnetic field before an eruption in AR NOAA 11726. They showed the formation of a pre-eruptive coronal flux rope and analyses the evolution of magnetic helicity and dynamics of the helicity eruptivity index. The flux rope constituted only a fraction of the whole active region. They noted that the index was steadily increasing when considering the whole domain, with no decrease after the eruption. When only taking into account the domain where the eruptive flux rope was located, the helicity eruptivity index displayed peaks before the eruption time. Analysing thoroughly the link between the variations of the helicity index and every form of activity developing in AR NOAA 11158, \citet{Green22} found the helicity ratio variations to be more pronounced during times of strong flux emergence, collision and reconnection between fields of different bipoles, shearing motions and reconfiguration of the corona through failed and successful eruptions. It was observed to a high degree that any form of eruptivity (jets, failed eruptions, eruptions) had a signature in the helicity eruptivity index. Even jets developing at a smaller scale than the whole active region, over which the helicity eruptivity index was calculated, were related with fluctuations of the index.

This motivates the present study to analyse the properties of helicities in coronal jet simulations, and the link between the generation of such type of activity with the helicity eruptivity index. In the present study, we perform new innovative analysis of the parametric 3D MHD simulations of \citet{Wyper18b} to investigate the time variations of magnetic energies and magnetic helicities. We analyse two simulations with a very similar set-up, one inducing a jet and one without eruptive activity. In both simulations helicity and energy are injected thanks to line-tied boundary forcing, although for a slightly longer time in the simulation in which a jet is induced. However the jet is not induced immediately after the forcing, but rather after a delayed period in which a reconfiguration of the magnetic system is observed. A period of less substantial reconfiguration is also noted in the stable configuration.  In the present work, we aim to compare the dynamics, in terms of energies and helicities, of this post-driving phase/reconfiguration phase in the Jet producing versus the Non-eruptive case. We also examine whether the transfer term between the two helicity components $\Hj$ and $\Hpj$ plays a major role in the helicity budgets as was observed in \citep{Linan18}. Finally, we want to see if the helicity eruptive index is able to discriminate the two simulations, the eruptive from the Non-eruptive one, and is able to provides sensible information about the eruptivity of the magnetic system.

Our manuscript is decomposed into different sections organised as follows. In \sect{Sim}, we first summarise the concept and properties of the numerical experiments of \citet{Wyper18b} that are analysed in the present study. In \sect{Method}, we then introduce the methods employed to estimate magnetic energy and helicity and their decomposition based on potential and non-potential magnetic field, as well as the helicity fluxes. The analysis of the dynamics of energies and helicities in the two simulations is presented in \sect{Results}. Finally, in \sect{Conclusion}, we summarise our results and discuss them in the broader context of the problematic of the trigger active solar events.     

\section{Non-eruptive and Jet producing numerical simulations} \label{sec:Sim}




\subsection{Numerical model} \label{sec:Simmodel}

Motivated by a growing number of jet observations revealing minifilament/sigmoid eruptions \citep[\eg][]{Raouafi10,Sterling15}, the jet simulations of \citet{Wyper17,Wyper18b} were designed to explore the nature of filament channel eruptions in coronal jets and how they compare to large-scale CME-producing active region eruptions. The key features of the model are that the initial magnetic field is comprised of a 3D magnetic null point topology above a bipolar surface flux distribution which is surrounded by uniform vertical (or tilted) open field. Line-tied surface motions lead to the formation of a filament channel at the centre of the bipole while maintaining the same surface flux distribution \citep{Pariat09a}. As outlined below, subject to sufficient forcing the filament channel becomes destabilised and erupts. This destabilisation is aided all or in-part by null point reconnection above the filament channel which as shown in \citet{Wyper17} is exactly analogous to the "breakout reconnection" hypothesis generating active region CMEs \citep{Antiochos99}.  \citet{Kumar18,Kumar19} amongst others have shown that this model captures many observational features of coronal jets. This realism along with the involvement of a flux rope in the eruption make this model an ideal test for the helicity index.

Here we focus on the simulation from \citet{Wyper18b} with vertical open field and consider two cases. The Jet producing simulation described in \citet{Wyper18b} in which the driving is ramped up to a constant speed over a period of $50$ non-dimensional time units, held constant until $t=300$ and then ramped down to zero (again over $50$ time units). And a new Non-eruptive case, similar to the first but where the driving is held constant instead until $t=250$ before being ramped down. Both simulations are identical, except the grid was allowed to adaptively refine one further level for the Jet producing case to better delineate the different phases of the eruptive evolution. However, as outlined below their early evolution prior to $t=250$ is quasi-identical. In both, the ideal compressible MHD equations are solved using the ARMS code \citep{Devore08}, with reconnection occurring due to diffusion intrinsic to the numerical scheme. For context, one time unit is roughly the Alfv\'{e}n travel time across the width of the separatrix dome based on the maximal Alfv\'{e}n speed on the surface.

\subsection{Common initial forcing phase} \label{sec:Simforcing}

\begin{figure*}
\centering
\includegraphics[width=\linewidth]{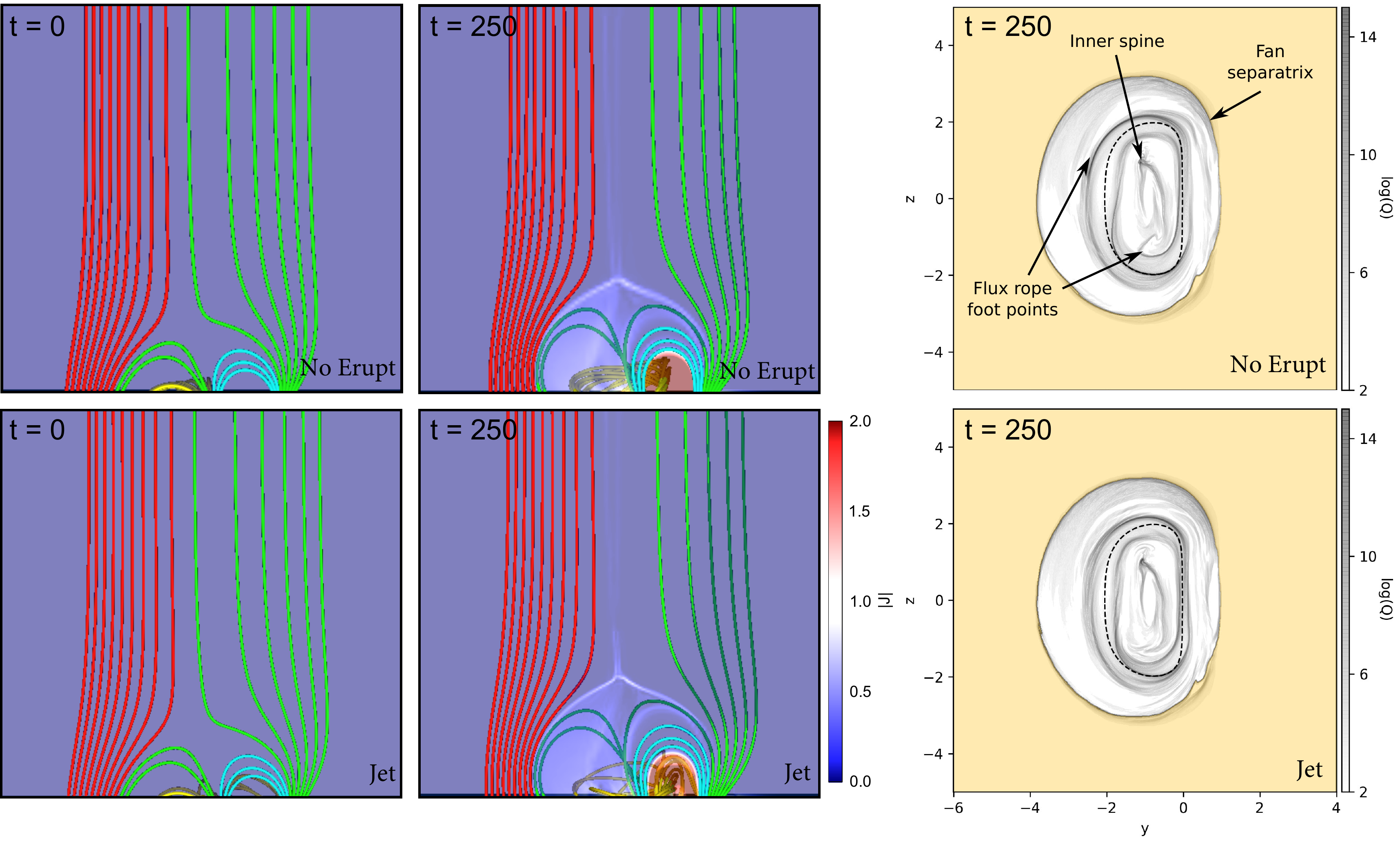}
\caption{
Left panels: current and field lines during the common initial phase $(0<t<250)$ of the simulations. Right: QSL distribution at $t=250$. The dashed line shows the PIL. Top panels: the Non-eruptive case. Bottom panels: the Jet producing case. Yellow shading indicates open field.}
\label{fig:JetSim_CommonPhase}
\end{figure*}

The left panels of \fig{JetSim_CommonPhase} show representative field lines and the current density in the two simulations at $t=0$ and at the end of the common driving phase ($t=250$). The cyan field lines connect the two halves of the surface bipolar patch. At $t=250$ these field lines form part of the strapping field above the filament channel formed by the action of the driving (yellow field lines). At the end of this common forcing phase the simulations are near identical. Only slight differences in the field line morphology within the filament channel are present by the end of this phase due to the differences in local resolution, with the better resolved jet case containing sheared field lines that extend slightly further along the polarity inversion line (PIL). 

The right panels of Figure \ref{fig:JetSim_CommonPhase} show the the squashing factor, Q, on the surface \citep{Titov02,Titov07,Pariat12}, with the yellow shaded region indicating the open field. The squashing factor is related to the gradients of the magnetic connectivity of the field lines. Volumes of high Q, named Quasi-Separatrix Layers \citep[QSLs,][]{Demoulin96a,Longcope05} delimit (quasi-)connectivity domains and represent preferential sites for the build-up of electric currents \citep{Aulanier05b,Aulanier06b}. True separatrices are always embedded in a QSLs halo \citep{Pontin16}, hence the Q distribution also captures the location of the fan and the spine of a 3D null-point \citep{Masson09,Masson17}.

Here, both distributions of Q are very similar, with the circular footprint of the fan separatrix and QSL around the inner spine in close agreement. Parallel strips of high Q flank the right side of the PIL (the centre of the surface bipole flux distribution) indicating that a small flux rope has formed 
as a result of gradients in the surface driving profile. This filament channel flux rope wraps around the polarity inversion line with foot points as indicated. One starts to observed, in particular for the inner flux rope footpoint, the characteristic hook shape in the distribution of Q associated with flux rope \citep{ZhaoJ16}.  

\subsection{Non-eruptive simulation} \label{sec:SimNonrupt}

\begin{figure}
\centering
\includegraphics[width=\linewidth]{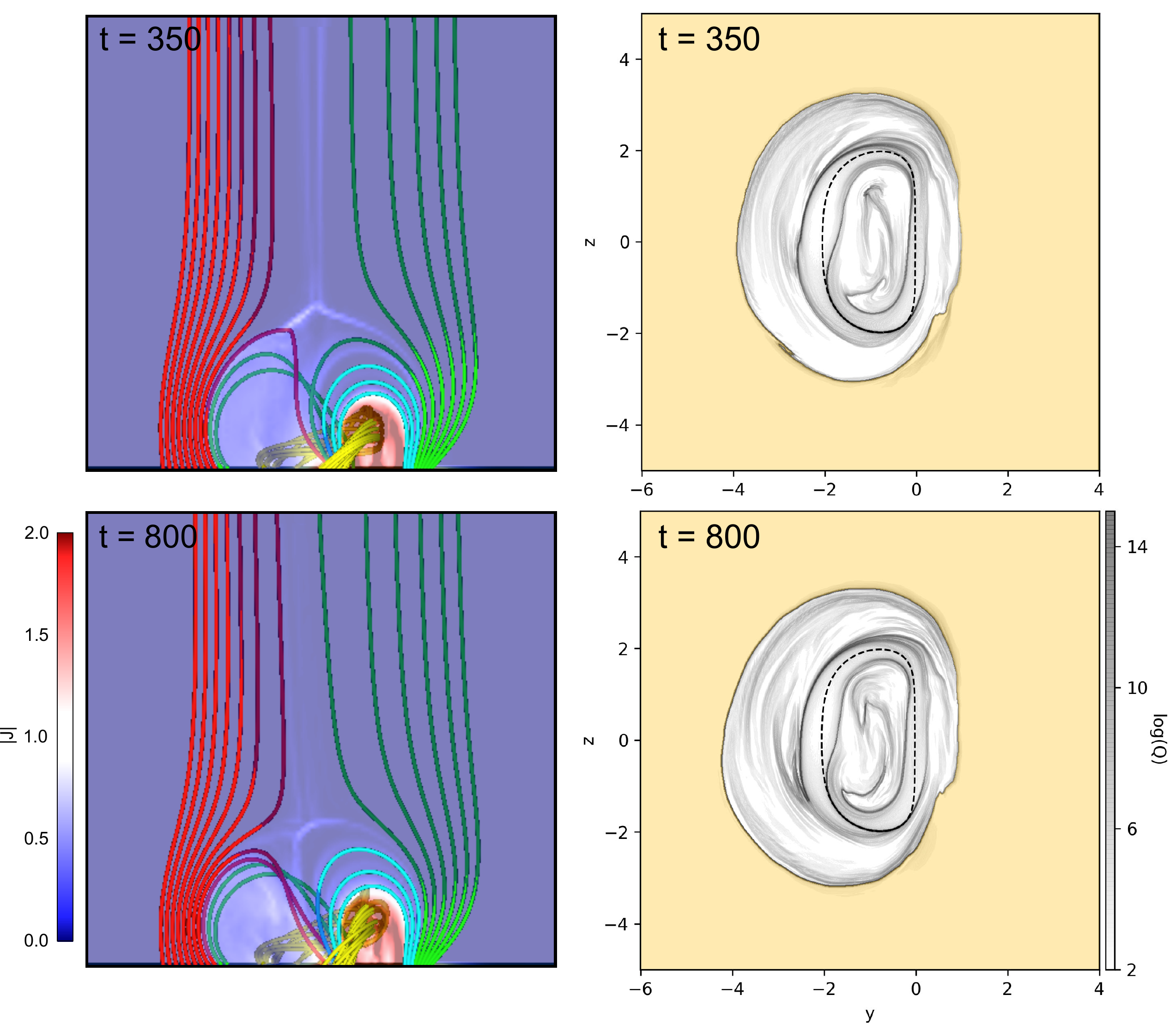}
\caption{Snapshots at t=350 \& 800, for the Non-eruptive simulation. Left Panels: current and field lines. Right panels: QSL distribution. Yellow shading indicates open field. 
}
\label{fig:JetSim_Stable}
\end{figure}

Beyond $t=250$ the driving in the Non-eruptive case ramps down to zero. This phase is named the post-driving phase of the Non-eruptive simulations. The injected shear sufficiently expands the closed field that the null point is stressed and low intensity reconnection is induced. \Fig{JetSim_Stable} shows the field lines and QSLs not long after the driving is halted and a substantial time later ($t=800$).  The low intensity reconnection has closed down some of the red open field lines while simultaneously opening up some of the strapping field (\fig{JetSim_Stable}, top right panel). This can also be seen in the leftward shift of the footprint of the fan separatrix (\fig{JetSim_Stable}, bottom right panel). By $t=800$ this low intensity reconnection has dissipated the stress around the null point and the reconnection effectively ceases while the filament channel remains stable. The system remains almost unchanging from then on.

\subsection{Jet producing simulation} \label{sec:SimJet}

\begin{figure*}
\centering
\includegraphics[height=0.9\textheight]{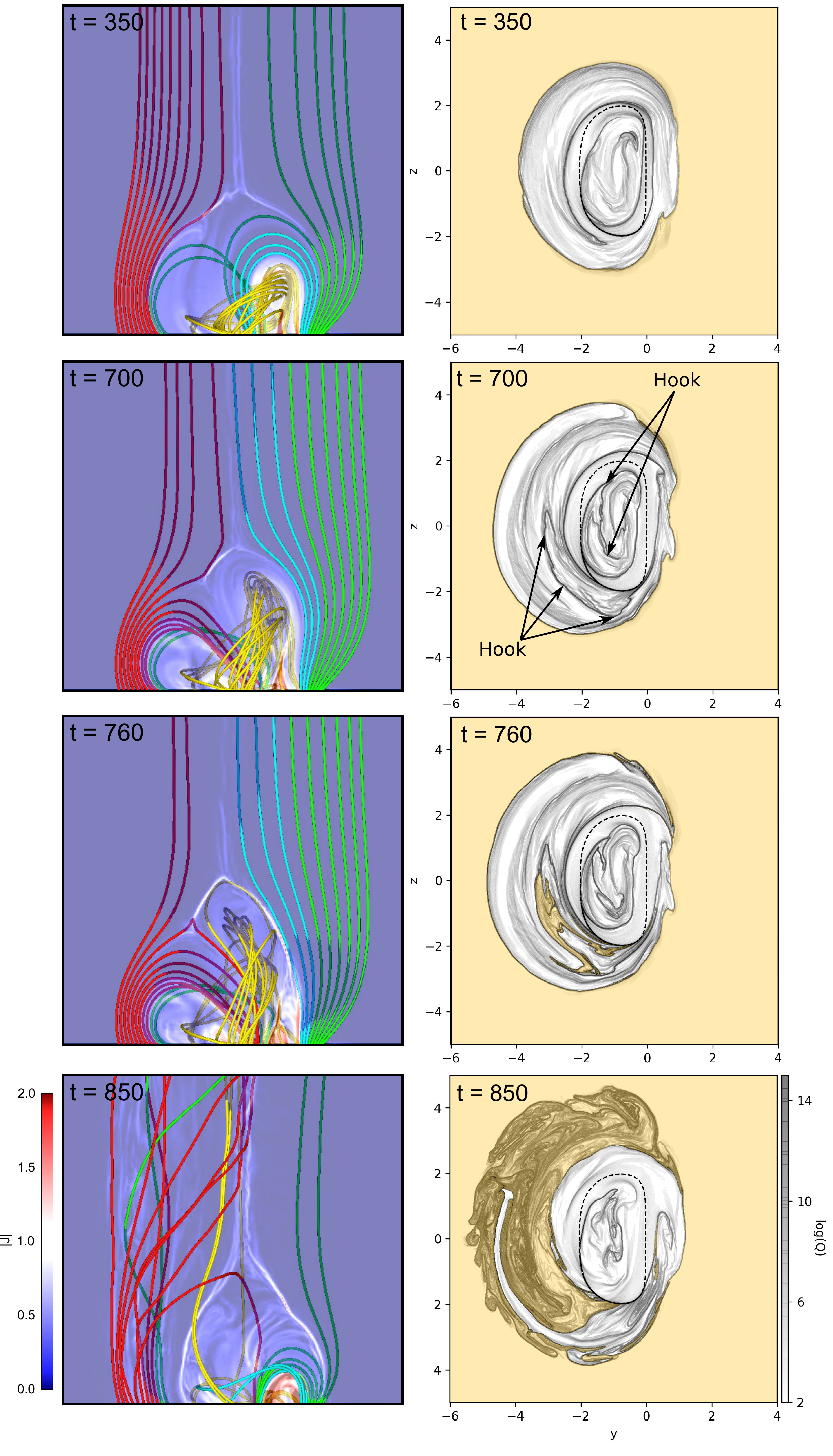}
\caption{Snapshots at t=350, 700, 760, \& 850 for the Jet producing simulation. Left panels: current and field lines. Right panels: QSL distribution. Yellow shading indicates open field.}
 \label{fig:JetSim_Jet}
\end{figure*}

By contrast in the Jet producing simulation the longer driving time tips the system into an unstable regime. This implies a point of no return is passed between when the driving is halted at $t = 300$ vs $t = 350$. In this case, after $t  = 350$, the system enters a long phase of sustained null point reconnection, denoted as the "breakout phase" in \citet{Wyper18b}, following a feedback between the upward expansion of the flux rope and the removal of strapping field above it (\fig{JetSim_Jet}, top left panels). At the same time reconnection also occurs at the 
current layer beneath the flux rope. The result is that the strapping cyan field lines are steadily removed from above the flux rope, while the flux rope itself both lengthens and increases in overall magnetic flux (compare the left panels at $t=350$ and $700$). That is to say during this phase a larger fraction of the closed field magnetic flux becomes part of a single, coherent flux rope, while simultaneously the strapping field linking with it is removed. The removal of strapping field is discernible in the squashing degree (Q) plot at $t=700$ by the leftward shift of the fan separatrix, while the broader area spanned by the QSL hooks indicates the increase in the magnetic flux contained within the flux rope. It should be noted that although the reconnection in both current sheets is sustained it is not explosive or impulsive during this phase and the flux rope rises slowly. This phase, between $t=350$ and $t\sim 740$, is labelled, in this study, the post-driving phase of the Jet producing simulation.

As more fully discussed in \citet{Wyper18b} an impulsive change in the evolution occurs when the strapping field is exhausted and the flux rope encounters the null point current sheet. This occurs around $t = 740$ after which the flux rope rapidly begins to reconnect with the open field, transferring a faction of the twist within the flux rope to the open field. This is shown in the QSL plot at $t=760$ by one foot point of the flux rope partly now residing in the open field region, while at $t = 850$ (once the jet is launched) the rest of the sheared closed field has now also become open. This transfer of twist in addition to the reconnection outflows are what form the jet \citep[\cf][]{Shibata86,Pariat09a,Pariat15a,Pariat16,Wyper17,Wyper18b}. This period is named the jet onset phase.

\section{Magnetic energies and helicities estimation methods} \label{sec:Method}

In the following section, we introduce the method used to numerically compute, in the two simulations, the magnetic energies and helicities, as well as some derived quantities such as helicity fluxes and the helicity eruptivity index, $\eta_H$. 

Our analyses primarily relies on the determination of the unique potential field $\vBp$ of $\vB$, which has the same flux distribution of $\vB$ through the boundary $\surf$ of the domain $\vol$ and satisfies:
\begin{equation}
\left\{\begin{array}{l}
\nabla\times \vBp =0 \\
\vn\cdot(\vB-\vBp)|_{\surf}=0
\end{array} \right.
\end{equation}
where $\vn$ is the outward-pointing unit vector locally normal to $\surf$. The potential field, $\vBp$, can thus be defined through the use of the scalar function, $\phi$, which is the solution of the Laplace equation with Neumann boundary conditions:
\begin{equation} \label{eq:Bp}
\left\{
\begin{array}{l} 
\vBp = \nabla\phi \\
\Delta\phi=0 \\
\left.\frac{\partial\phi}{\partial n}\right|_{\mathrm{S}} = (\textbf{n}\cdot\textbf{B})|_{\mathrm{S}}.
\end{array} \right.
\end{equation}
For a given magnetic field, $\vB$ studied in a simply connected domain, the potential field $\vBp$ is uniquely defined. The magnetic field $\vB$ is thus uniquely decomposed as:
\begin{equation}
\vB= \vBp +\vBj \,,
  \label{eq:Bdec}
\end{equation}
with $\vBj$ the non-potential field, uniquely defined as the difference $\vBj=\vB-\vBp$. The field $\vBj$ is the current-carrying part of the field since $\nabla\times \vB = \nabla\times \vBj = \mu_0 \vj$, following the Amp\`ere–Maxwell law with $\vj$ the electric current density and $\mu_0$ the magnetic constant.

\subsection{Magnetic Energy decomposition}
\label{sec:methodenergy}

Using the decomposition of $\vB$ into current-carrying and potential components (\cf \eq{Bdec}), for a strictly solenoidal field ($\divB=0$), the total magnetic energy $\E$ can be classically decomposed as (Thompson's theorem):
\begin{equation}
\E= \Ep +\Ej \,,
  \label{eq:thomson}
\end{equation}
with $\Ep$ is the potential energy and $\Ej$ is the energy of the non-potential field, frequently also called the free magnetic energy.

When $\vB$ is not strictly solenoidal, for example when $\vB$ is represented over a discrete mesh such as in numerical experiments, \citet{Valori13} has shown that the energy of the magnetic field in $\vol$ can be distributed into solenoidal and nonsolenoidal contributions, as in:
\begin{equation}
\E= \Ep +\Ej +\EdivBp +\EdivBJ +\Emix \,,
  \label{eq:thomson-valori}
\end{equation}
where $\Ep$ and $\Ej$  are the energies associated to the potential and current-carrying solenoidal contributions,
$\EdivBp$ and $\EdivBJ$  are those of the nonsolenoidal contributions, and   $\Emix$ is a nonsolenoidal mixed term \citep[see Eqs.~(7,8) in][for the corresponding expressions]{Valori13}.
All terms in \eq{thomson-valori} are positively defined, except for $\Emix$.
For a perfectly solenoidal field, $\EdivBp=\EdivBJ=\Emix=0$, that is the Thomson's theorem is recovered.

Following \citet{Valori16}, to analyse the eventual impact of the non-solenoidality in the discretised data, we consider a single number for characterizing the energy associated to nonsolenoidal components of the field, given by: 
\BE
  \Ediv=\EdivBp +\EdivBJ +|\Emix| \,.
  \label{eq:ediv}
\EE
This method, which has now been regularly used \citep{Valori16,Pariat17,Moraitis19b,Thalmann19a,Thalmann19b,Thalmann21}, is basically a numerical verification of Thomson's theorem, and allows one to quantify the effect of a (numerical) finite divergence of the magnetic field in terms of associated energies. The derived values of $\Ediv$ in both simulations are extremely small and only corresponds to about $0.1-0.2 \%$ of $\E$. These values can be compared to the different test cases of \citet{Valori13}, with similar amplitudes to the analytical test over a discrete grid. The simulations are thus highly solenoideal. For these values of $\Ediv/\E$, magnetic helicity estimations are extremely reliable \citep[\cf Sect. 7 of][]{Valori16}.

\subsection{Relative magnetic helicity decomposition}
\label{sec:methodhelicity}

In the fixed volume $\vol$ bounded by the surface $\surf$, the magnetic helicity, $\Hm$, is classically defined as:
\begin{equation} \label{eq:ClassicH}
\Hm = \intv \vA\cdot\vB \dV,
\end{equation}
with $\vA$ the vector potential of the studied magnetic field $\vB$, i.e $\curlA=\vB$. In practice, this scalar description of the geometrical properties of magnetic field lines is relevant only if the magnetic field is tangential to the surface, i.e $V$ is a magnetically bounded volume. Indeed, the magnetic helicity is gauge invariant if and only if this condition is respected. For the study of natural plasmas, especially in the solar physics, the magnetic field does not satisfied this condition, the solar photosphere being subject to significant flux.

In order to lift this caveat, \citet{BergerField84} introduced the concept of relative magnetic helicity, a gauge invariant quantity, based on a reference field. Using $\vAp$ the vector potential of the potential field, $\vBp=\curlAp$, the relative magnetic helicity provided by \citet{Finn85} is:
\begin{equation}
\Hv= \intv (\vA+\vAp)\cdot(\vB-\vBp) \, \dV. \label{eq:h}
\end{equation}
In this form, the relative magnetic helicity is gauge invariant for both $\vAp$ and $\vA$. The difference between the potential field and the magnetic field can be written as a non-potential magnetic field, $\vBj=\vB-\vBp$, associated with the vector $\vAj$, defined as $\vAj=\vA-\vAp$, such as $\curl \vAj=\vBj$. Following \citet{Berger03}, $\Hv$ can be divided into two gauge invariant quantities \citep[see also][]{Pariat17,Linan18,Linan20}:
\begin{eqnarray} \label{eq:hvdec}
 \Hv&=&\Hj+\Hpj,\\
 \Hj&=&\intv \vAj \cdot \vBj \dV, \label{eq:hj}\\ 
\Hpj&=&2\intv \vAp \cdot \vBj \dV, \label{eq:hpj}
\end{eqnarray}
where $\Hj$ is the non-potential magnetic helicity associated to the current-carrying component of the magnetic field, $\vBj$, and $\Hpj$ is the volume-threading helicity involving both $\vB$ and $\vBp$. By construction, both $\Hj$ and $\Hpj$ are gauge invariant, since $\vBj$ has no normal contribution to the surface $\surf$.

\subsection{Helicity eruptivity index}
\label{sec:methoderuptivityindex}

Following \citet{Pariat17}, we define the helicity eruptivity index, $\eta_H$, as the ratio of the non-potential helicity to the total relative helicity:
\begin{equation}
\eta_H= \frac{|\Hj|}{|\Hv|} \, . \label{eq:etah}
\end{equation}
This non-dimensional ratio is here defined positively. It shall be noted that since helicities are signed quantities, $\Hj$ and $\Hv$ can have opposite signs. The index $\eta_H$ is also not bounded by $1$ since $\Hj$ can exceed $\Hv$. This may happens in the case where $\Hpj$ and $\Hj$ have opposite signs, as in the jet case analysed by \citet{Linan18}. In the present simulations however, all helicities are positive and one thus have: $\eta_H = \Hj/\Hv$.

Let us note that \citet{YangK20} has proposed an alternative definition of the helicity eruptivity index, based on a periodic potential field. This index may be more suited for systems with a higher degree of periodicity, very distinct from the one studied here. 

\subsection{$\Hj$ and $\Hpj$ time variations}
\label{sec:methodhelicityfluxes}

The study of the time variations of relative magnetic helicity has now benefited from two decades of investigation \citep[\eg][]{Chae01,Chae07,Pariat05,Pariat15b,Dalmasse14,Schuck19}. Relative magnetic helicity being a conserved quantity in ideal MHD, its time variations can be solely written as the results of a flux through the boundary of the studied domain \citep[\cf Sect. 2 of][]{Pariat15b}. In ideal MHD, there is no volume term that would dissipate/create magnetic helicity.
Additionally, the time variations of $\Hv$, $\dHdt$ can trivially be related to the time variations $\dHjdt$ and $\dHpjdt$ of $\Hj$ and $\Hpj$ respectively:
\begin{equation}
\fdHdt = \fdHjdt + \fdHpjdt \, . \label{eq:dhdt}
\end{equation}

Motivated by the interest to understand the properties of $\Hj$ and $\Hpj$, \citet{Linan18} have studied the time variation of these helicities. \citet{Linan18} have established the following gauge invariant equations of the evolution equations of $\dHjdt$ and $\dHpjdt$:
\begin{eqnarray}
\fdHjdt &=& \fdHjdtdiss + \fdHjdtown + \fdHjdttrans \label{eq:dhjdtni} \\
\fdHpjdt &=& \fdHpjdtdiss  + \fdHpjdtown + \fdHpjdttrans \label{eq:dhpjdtni} 
\end{eqnarray}
The terms $\dHjdtdiss$ and $\dHpjdtdiss$ \citep[which formulations can be obtain respectively in Eqs. (49) \& (54) of][]{Linan18} are volume dissipation terms. These terms are null in ideal MHD. The terms $\dHjdtown$ and $\dHpjdtown$ are variations terms which are proper to $\Hj$ and $\Hpj$ respectively. They are the sum of diverse terms and their complete formulation can respectively be found in Eqs.(51) and (55) of \citet{Linan18}. In a specific set of gauges (the coulomb gauges), $\dHjdtown$ and $\dHpjdtown$ can be expressed solely as terms of fluxes. Hence, $\dHjdtown$ (resp. $\dHpjdtown$) corresponds to the injection/expulsion of $\Hj$ (resp. $\Hpj$) through the boundary $\surf$ of $\vol$. Finally, $\dHjdttrans$ and $\dHpjdttrans$ are volume terms with equations given by:
\begin{equation}
\fdHjdttrans = - \fdHpjdttrans = -2 \intv  (\vv \times \vB) \cdot \vBp \dV \, . \label{eq:trans}
\end{equation}
These volume terms have opposite sign : they correspond to terms of transfer of helicity between $\Hj$ and $\Hpj$. \citet{Linan18} have thus found that unlike magnetic helicity, $\Hj$ and $\Hpj$ are not conserved quantities and have highlighted the existence of a gauge-invariant volume term that acts to convert $\Hj$ into $\Hpj$ and inversely.

In ideal MHD, the dissipation terms are strictly null. Even when non-ideal effects such as magnetic reconnection is present, the dissipation of magnetic helicity is thought to be very limited \citep{Berger84b,Pariat15b}. Similarly to the simulation of \citet{Pariat09a}, the simulations studied here are modelled with the ARMS solver without explicit resistivity but with an adaptive mesh refinement strategy which increases the resolution at current sheets, where magnetic dissipation is the largest. Analysing the jet simulation of \citet{Pariat09a}, \citet{Pariat15b} demonstrated that the dissipation of relative magnetic helicity, $\Hv$ was extremely limited (below $2\%)$, even when intense magnetic reconnections / reconfiguration of the system was ongoing. Following \citet{Linan18}, we verified that the dissipation of $\Hj$ and $\Hpj$ was also very limited in the presently studied simulations. In such case, the evolution equations of $\Hj$ and $\Hpj$ can thus be limited to:
\begin{eqnarray}
\fdHjdt &=& \fdHjdtown + \fdHjdttrans  \, , \label{eq:dhjdt} \\
\fdHpjdt &=& \fdHpjdtown + \fdHpjdttrans \, . \label{eq:dhpjdt} 
\end{eqnarray}

Following \citet[][\cf Sect. 3.4]{Linan18}, we assessed the validity of the assumption of near ideality. We measured the difference between the time derivative of $\Hj$ and $\Hpj$ with the direct estimation of $\dHjdt$ and $\dHpjdt$ in both simulations. We found that the relative error was at most $7\%$, which remains very small.  This is in the range of what was obtained in \citet{Linan18} for the different MHD simulations analysed.  The main differences occur during the period of strong evolution of $\dHjdt$ and $\dHpjdt$ and thus the difference likely mainly results from the relatively low cadence of the data which does not permit to evaluate optimally the time derivative of $\Hj$ and $\Hpj$. We are overall confident that the dissipation of helicities remains negligible in comparison to the other terms.   

\subsection{Methods to estimate energies and helicities}
\label{sec:methodestimation}

In order to compute the different helicities and energies at each time in the simulation, we follow the procedure of \citet{Valori12}, \citet{Valori13} and \citet{Linan18}. 

We focus our analysis on data cubes of $\vB$ and $\vv$ extracted from the adaptive mesh grid of each simulation. The data cubes were extracted on a regular grid in a sub-volume with $x\in [0,10.8]$ and $y$ and $z\in\pm 5.8$. In both simulations the grid within this volume has a fixed minimum of $4$ levels of refinement (see Fig. 3 in \citet{Wyper18b}). The regular grid for the data cubes is coincident with this uniform local grid. In the Jet producing case this leads to a slight coarsening of the grid in places where the grid adaptively refined to one level higher. As they are integral quantities, and as shown in \citep{Pariat15b} this has a negligible effect on the helicities. .

The time-sequence of datacubes of the magnetic field $\vB$ permits to compute all the magnetic energies and magnetic helicities of  \eqs{thomson-valori}{hvdec}. First, the scalar potential $\phi$ is obtained from a numerical solution of the Laplace equation (\cf. \eq{Bp}). The solenoidal potential field $\vBp$ and the solenoidal non-potential field $\vBj$ are derived following \citet[][ see Sects. 3.1 \& 3.2]{Valori13}. These fields permits to derive the different energies of \eq{thomson-valori} and in particular $\E$, $\Ej$, $\Ep$ and $\Ediv$ (\eq{ediv}).

In order to compute the helicities, the potential vectors $\vA$ and $\vAp$ are then estimated using the DeVore-Coulomb gauge defined in \citet{Pariat15b}, based on Eq. (14) of \citet{Valori12}. Given that the system is a solar-like active region, with more intense magnetic field at the bottom boundary, following previous practice, the 1D integration involved is started from the top of the domain in order to minimise errors \citep[\cf discussions in][]{Pariat15b,Pariat17}. The gauge used to compute the potential vectors is fully fixed. From the derived potential vectors, we obtain the helicities, $\Hv$, $\Hj$ and $\Hpj$ from \eqss{hvdec}{hpj}.  As a sanity check, we also performed the computation in a different gauge \citep[see \eg][for other gauge choices]{Pariat17}. Given the low solenoidality of the magnetic field (low $\Ediv$, \cf \sect{methodenergy}), we found no noticeable difference between the computation performed in the different gauges.

The computation of the time variations of $\Hj$ and $\Hpj$ can be then done independently from the estimations of the volume helicities \citep{Linan18,Linan20}. In addition of the knowledge of the magnetic fields ($\vB$, $\vBp$ and $\vBj$) and from the estimation of their vector potential ($\vA$, $\vAp$ and $\vAj$), the estimation of the terms of \eqs{dhjdt}{dhpjdt} requires the knowledge of the datacubes of the plasma-velocity field $\vv$, which is extracted from the simulation similarly to $\vB$. This allows us to determine, $\dHjdtown$, $\dHpjdtown$, $\dHpjdttrans$, $\dHjdttrans$ and their sum $\dHjdt$ and $\dHpjdt$.

All these quantities are computed in both simulations, at each time step. This permit to analyse finely the dynamics of the magnetic energies and of the helicities in the Jet producing simulation and compare it with the Non-eruptive one.

\section{Magnetic energies and helicities dynamics in the simulations} \label{sec:Results}

In this section, we describe the evolution, in both the Non-eruptive and the Jet producing simulation, of the different magnetic energies and helicities, as determined by the methods described in \sect{methodestimation}. The focus being on the pre-eruptive phase , the description of the energies and helicity evolution during the jet (for the Jet producing simulation) will only be overviewed, without going into details.

\subsection{Magnetic energies evolution} \label{sec:Edyn}

The evolution of the total magnetic energy, $\E$, of the potential energy $\Ep$ and of the free magnetic energy, $\Ej$ is presented in the top panel of \fig{EnergyHelicity} and the values at a few selected times are given in \tab{EHvalues}. Because of their very low value, thanks to the excellent solenoidality of the $\vB$ (\cf \sect{methodenergy}), the solenoidal terms entering in the decomposition of $\E$ (\cf. \eq{thomson-valori}) are not represented.

Per design the field is initially potential, and one has  $\E(t=0)=\Ep(t=0)$ and $\Ej(t=0)=0$. Thanks to the bottom boundary driving motions, free magnetic energy is injected in the system. During the common driving phase, $\Ej$ monotonically (and almost linearly) increases in both simulations. Meanwhile, $\Ep$ very slightly decreases, with $(\Ep(t=300)-\Ep(t=0))/\Ep(t=0)\sim0.02$. By design of the driving pattern in the simulations, the vertical component of $\vB$ is kept fixed. One could believe that $\Ep$ would remain constant. However, the forcing enhance the transverse field in the close field domain. Because of the increase of the magnetic pressure, the closed-field domain bulges pushing slightly the open field. The distribution of the normal component to the side boundaries of the system are thus slightly changing inducing the observed small evolution of $\Ep$. This variation is however very small compared to the injection of $\Ej$, and $\E$ therefore steadily increases during the driving phase. 

At $t=300$, at the end of the forcing for the Non-eruptive simulation, $\Ej$ represents 34\% of $\E$ (\cf. top panel of \fig{RelativeFractions}). At the same instant,  $\Ej/\E \sim 37 \%$ for the Jet producing simulation. This value is slightly larger for the Jet producing case because the driving motion has been ramped down earlier for the Non-eruptive run.

During the post-driving phase of the Non-eruptive simulation, $\Ep$ remains basically constant (within $0.3\%$). The non-potential energy $\Ej$ very slightly decreases (see top panel of \fig{EnergyHelicity}). The relative decreases by the end of the Non-eruptive simulation is of the order of 3-4\%.
This decrease is likely due the the low intensity reconnection and the mild reconfiguration taking place in the system during that phase. Accordingly, $\E$ also decreases but this only corresponds to about 1\% of relative variation. As can be noted in the top panel of \fig{RelativeFractions}, the ratio of the free energy to the total energy remains constant during this post-driving phase for the Non-eruptive simulation.

The Jet producing simulation is driven until $t=350$ and thus benefits from a larger energy input. The peak value of $\Ej$ is about 26\% higher for the Jet producing simulation than for the Non-eruptive simulation. At $t=360$ just after the end of the forcing, the ratio of the free energy normalised by the total energy has reached $0.39$, which is about 15\% higher than the maximum ratio of the Non-eruptive simulation.

As with the Non-eruptive simulation, the reconnection and the reconfiguration occurring during the post-driving phase induces a decrease of $\Ej$ and $\E$ (while $\Ep$ stays almost constant). However, since the reconnection dynamics has a stronger intensity in the Jet producing simulation, the decrease is more marked in absolute value: $\E$ and $\Ej$ decreases by about 30 energy units between $t=360$ and $t=740$. However, since the Jet producing simulation had a larger free energy content, in relative value,  $\Ej$ and $\E$ respectively decreases by 4\% and 1.6\%. This relative variation is thus very similar to the energy change observed during the post-driving phase of the Non-eruptive simulation. In term of energy, the reconfiguration in the post-driving phase relatively impacts the system in a similar way.

Finally, after $t\simeq 750$, the onset of the jet is characterised by a sudden decrease of $\Ej$, as magnetic energy is dissipated and partly converted to kinetic energy.  $\Ep$ displays only weak variations, due to the small change of the magnetic flux distribution on the side and top boundaries.

\subsection{Magnetic helicities evolution} \label{sec:Hdyn}

The evolution of the total magnetic helicity, $\Hv$, of the non-potential helicity $\Hj$ and of volume-threading helicity, $\Hpj$ is presented in the bottom panel of \fig{EnergyHelicity} and their values at a few selected times are given in \tab{EHvalues}.

The initial configuration being potential, the system is void of helicity and the three helicities are null at $t=0$. During the common driving phase, shear and twist being injected, the total helicity monotonically increases. Unlike with magnetic energy, for which the increase was directly due to the injection of free magnetic energy, $\Ej$, the helicity injection presents three phases. First, between $t=0$ and $t=75$, $\Hv$ grows mainly due to the increase of $\Hpj$, while $\Hj$ very mildly increases. Then between $t=75$ and $t=150$, $\Hj$ starts to increase and $\Hv$ grows thanks to the increase of both $\Hj$ and $\Hpj$. The growth of $\Hpj$ is however becoming weaker and weaker, eventually reaching a maximum around $t\sim200$ and even decreasing. Hence, between $t=150$ and $t=300$ the increase of $\Hv$ is primarily due to $\Hj$.
It is worth noticing that the same dynamics of helicities were noted for the jet simulation analysed in \citet[][see Fig. 3]{Linan18}: $\Hv$ was growing first thanks to $\Hpj$, which eventually later decreased while $\Hj$ became the dominant contributor to $\Hv$. The analysis of the helicity fluxes, detailed in \sect{Hflux}, permits to better understand this evolution.  

For the Non-eruptive simulation, during the post driving phase, the dynamics of the helicities is in agreement with the evolution of the energies. All three helicities very weakly decrease: between $t\sim350$ and $t\sim900$, $\Hv$, $\Hj$, and $\Hpj$ display a relative variation lower than 3\%. This decrease is in line with the variation of $\Ej$ (and $\E$) observed in the same period, and likely due to the weak intensity reconnections occurring then.

On the contrary, the helicities in the Jet producing simulation presents sensible variations which were not observed with the energies during the post-driving phase. Even though the bottom boundary forcing has been halted, one observes a further decrease of $\Hpj$ and increase of $\Hj$. Between $t=360$ and $t=700$, $\Hj$ has a relative increase of 4\%. This increase, while not as strong as during the driving phase, is relatively constant and is strikingly in opposition to the observed decrease of $\Ej$ during the same period. The dynamics of $\Hv$ is however mostly dominated by the decrease of $\Hpj$ during this phase. While $\Hpj$ is roughly constant between $t=360$ and $t=460$, one observes a strong constant decrease between $t=460$ and $t=740$: $\Hpj$ presents a relative variation of 27\%. $\Hv$ thus similarly decreases. 
This evolution is present while no external forcing is applied to the system. The origin of this evolution is likely related to the important magnetic reconfiguration observed to occur within the magnetic system of the Jet producing simulation. During this phase, the Jet producing simulation witnesses both a more intense and longer current sheet at the null point, with more reconnection allowing strapping closed field lines to open, and simultaneously a more intense current sheet beneath the flux rope inducing both a strengthening in flux of the flux rope and its rise. In the present numerical experiment, helicities, as global scalar quantities cannot discriminate which dynamics (if not both) are responsible for the decrease of $\Hpj$. In any case the magnetic helicities thus appears to be much more sensitive to the magnetic reconfiguration observed in the system than the magnetic energies. $\Hpj$ presents a dynamic which is even more strongly marked as the system gets closer to the jet generation phase.

After $t=740$, the helicities dynamics in the Jet producing simulation is evidently marked by the eruptive process. Similarly to $\Ej$, $\Hj$ decreases strongly. Meanwhile $\Hpj$  first markedly increases and then decreases. $\Hv$ is dominated by the strong decrease of $\Hj$ and also diminishes. 

Overall, while the driving phase shows similarities between the energies and helicities dynamics, the post-driving phase displays very distinct behaviors. While energies do not display significant evolution, both for the eruptive and the Non-eruptive simulations, the helicities clearly discriminates the two numerical experiments. While the Non-eruptive simulation does not display significant changes during the post-driving phase, the Jet producing simulation is marked by variations of the helicities. Hence, unlike the energies, the helicities are able to capture the reconfiguration of the system which is occurring in the post-driving phase of the Jet producing simulation. The helicities are thus able to uniquely capture key dynamics of the magnetic system to which the energies are blind.

\subsection{$\Hpj$ and $\Hj$ conversion} \label{sec:Hflux}

The analyse of the time variations of $\Hj$ and $\Hpj$ permits to better understand the dynamics of helicity in the simulations. \Fig{FluxesHjHpj} presents the different terms of \eqs{dhjdt}{dhpjdt} for each simulation.

\subsubsection{Driving phase}

Starting with the evolution of $\dHjdt$ during the driving phase of Non-eruptive simulation (upper left panel of \fig{FluxesHjHpj}), one sees that the initial increase of $\Hpj$ results first from $\dHpjdtown$, that is from the injection of $\Hpj$ thanks to the boundary forcing motions. The curve of $\dHpjdtown$, follows the boundary driver, with first an increase between $t=0$ and $t=50$ as the boundary motions are ramped up, then a constant intensity before being ramped down between $t=250$ and $t=300$. While $\dHjdttrans$ is initially null until $t\sim50$, it then presents increasing negative values until $t=250$. It means that $\Hpj$ is being converted into $\Hj$. As a consequence, one observes \fig{EnergyHelicity} (lower panel) that $\Hpj$ first increases (dashed blue line). While the injection of $\Hpj$ is initially dominant, as $\dHjdttrans$ is becoming more and more intense, $\dHpjdt$ is becoming weaker and weaker. The increase of $\Hpj$ is thus being reduced, as is noted in  the lower panel of \fig{EnergyHelicity} (dashed blue line), reaching a maximum near $t\sim 200$. For a short period, around $t\sim250$, $\dHjdttrans$ is even becoming dominant over $\dHpjdtown$ (\cf \fig{FluxesHjHpj}) : $\Hpj$ is transferred faster into $\Hj$ than its injection by the boundary motion: the curve of $\Hpj$ (\cf \fig{EnergyHelicity}, lower panel) thus slightly decreases. 

The time evolution of $\Hj$ during the driving phase of the Non-eruptive simulation is very different from the one of $\Hpj$ (see lower left panel of \fig{FluxesHjHpj}). There is basically no injection of $\Hj$ thanks to the boundary driving motions: $\dHjdtown$ is constantly null. The variations of $\dHjdt$ is exclusively due to $\dHjdttrans$, meaning that $\Hj$ is uniquely formed thanks to the conversion from $\Hpj$. Since $\dHjdttrans$ is regularly increasing (having the opposite sign of $\dHpjdttrans$), $\Hj$ rapidly increases, as observed in the lower panel of \fig{EnergyHelicity} (dashed red line), although the rise of $\Hj$ is delayed compared to $\Hpj$. As the boundary driving motions are ramped down, the conversion of $\Hpj$ stops and the increase of $\Hj$ is drastically reduced.

For the Non-eruptive simulations, for $t>300$, during the post driving phase, all helicity variations terms are very small in comparison to the driving phase (left panels of  \fig{FluxesHjHpj}). They are close to zero, although not completely null as will be discussed later. Henceforth, $\Hj$ and $\Hpj$ remain almost constant after $t=300$ for this Non-eruptive case.

The helicity dynamics for the Jet producing simulation is completely equivalent to the Non-eruptive one during the driving phase (\cf right panels of \fig{FluxesHjHpj}). The curves of $\dHpjdt$, $\dHjdt$ and their decomposition present the same overall shape and intensity. The main difference between the two simulations during this driving phase is the longer driving time.  The primary source of helicity comes from $\dHpjdtown$ which generates an increase of $\Hpj$ (initial positive values of $\dHpjdt$). However, $\Hpj$ is converted into $\Hj$ and this conversion process eventually dominates $\dHpjdt$, which becomes negative. $\Hpj$ thus decreases. Because of the longer driving, the decrease of $\Hpj$ is more marked in the Jet producing simulation compared to the Non-eruptive one (see lower panel of \fig{EnergyHelicity}, continuous blue line). The non-potential helicity $\Hj$ also does not present proper injection ($\dHjdtown$ is almost null) and $\Hj$ is exclusively formed by conversion from $\Hpj$. Thanks to the longer driving time in the Jet producing simulation, $\Hj$ benefits from a longer time for conversion from $\Hpj$ and can thus reach larger values than in the Non-eruptive case (\fig{EnergyHelicity}, lower panel, continuous red line). This dynamics is fully consistent with the results of the analyse of the helicity dynamics of a jet simulation by \citet[][\cf Figure 11]{Linan18}.

\subsubsection{Post driving phase}
While the dynamics of $\Hj$ and $\Hpj$ are very similar for both simulations during the driving phase, strong differences appear between the two cases during the post driving phase, between $t=350$ and $t=700$. In order to better see the time variations of $\Hj$ and $\Hpj$, \fig{FluxesHjHpjzoom} presents a zoomed view of the evolution of $\dHpjdt$ and $\dHjdt$ during the post driving phase for each simulation. Three main differences can be noted between the Jet producing case and the Non-eruptive simulation: $\dHjdttrans$ (and $\dHpjdttrans$) has an opposite sign in the two simulations, its intensity is about twice larger for the Jet producing simulation, and $\dHpjdt$ is also significantly higher in the Jet producing case.

For the Non-eruptive simulation, during the post-driving phase, $\dHpjdttrans$ is constantly positive with an intensity lower than $0.1$ (\cf upper left panel of \fig{FluxesHjHpjzoom}). It implies a small conversion of $\Hj$ into $\Hpj$. Meanwhile $\dHpjdtown$ oscillates and is in average slightly negative. This corresponds to a small ejection of $\Hpj$ through the side boundaries while the magnetic system is slowly reconfiguring. As a results, $\dHpjdt$ oscillates around zero and hence $\Hpj$ is constant. Since $\dHjdtown$ is almost null (\fig{FluxesHjHpjzoom}, lower left panel), similarly to the driving phase, $\dHjdt$ is equal to $\dHjdttrans$ (\ie $-\dHpjdttrans$), meaning a slow transfer of $\Hj$ into $\Hpj$. This conversion is sufficiently small as to be barely discernible in the curve of $\Hj$ during the post driving phase of the non-eruptive simulation  (\fig{EnergyHelicity}, lower panel, dashed red line).

The time variations of $\Hj$ and $\Hpj$ are very different for the Jet producing simulation. Instead of being positive, $\dHpjdttrans$ is negative during the post-driving phase of the Jet producing simulation (\cf upper right panel of \fig{FluxesHjHpjzoom}). Respectively, instead of being negative in the Non-eruptive case, $\dHjdttrans$ is here positive (\fig{FluxesHjHpjzoom}, lower right panel). Similarly to the Non-eruptive simulation, $\dHjdtown$ is almost null and $\dHpjdtown$ is overall negative. As in the Non-eruptive simulation, there is no proper injection of $\Hj$ and $\Hpj$ is being ejected from the system though the side boundaries. However, the intensity of $\dHpjdtown$ is about twice larger in the Jet producing case compare to the Non-eruptive case (\fig{FluxesHjHpjzoom}, top panels). Contrary to the Non-eruptive case, since $\dHpjdtown$ and $\dHpjdttrans$ have the same negative sign for the Jet producing case, $\dHpjdt$ is markedly negative which corresponds to a sensible decrease of $\Hpj$ during this post-driving phase (\cf continuous blue line in the lower panel of \fig{EnergyHelicity}).

The intensity of  $\dHjdttrans$  is about $0.2$ for the Jet producing simulation, which is about twice the intensity in the Non-eruptive case (\fig{FluxesHjHpjzoom}, bottom panels).
Rather than a conversion of $\Hj$ into $\Hpj$, the post driving phase is marked by a further conversion of $\Hpj$ into $\Hj$. The conversion that was already ongoing during the driving phase keeps on, although at a smaller rate.  In the post driving phase of the Jet producing simulation $\Hj$ is thus further rising (\cf \fig{EnergyHelicity}, lower panel, continuous red line). 

The reconfiguration of the magnetic system which is observed during the post-driving phase of the Jet producing simulation (\cf \sect{SimJet}) is thus fundamentally different from the one happening in the Non-eruptive simulation. While in the Non-eruptive case the reconfiguration induce a minor decrease of $\Hj$ which is transformed in $\Hpj$ that is ejected out of the domain, in the eruptive simulation $\Hpj$ is partly ejected and partly transformed into $\Hj$. The evolution induced simultaneously by the intense null point reconnection, the reconnection beneath the flux rope and the rise of the flux rope, impact the helicities distribution of the Jet producing simulation, without here being possible to causally link each system dynamics to a specific helicity evolution.
The non-potential helicity $\Hj$ is thus rising while $\Hpj$ decreases: this naturaly leads to an evolution of the helicity eruptivity index $\eta_H$, as will be discussed in \sect{Hindex}.

Finally, during the jet generation phase of the Jet producing simulation (\ie for $t>750$), $\Hpj$ and $\Hj$ present strong variations (\cf right panels of \fig{FluxesHjHpj}). The evolution during that phase is completely similar to the jet simulation of \citet{Pariat09a} that has been analysed in \citet[][\cf Figure 11]{Linan18}. $\Hj$ first and mainly decreases because it is converted into $\Hpj$: $\dHjdttrans$ presents a strong negative peak. $\Hj$ thus increases (positive $\dHpjdt$) thanks to a positive $\dHpjdttrans$. However the increase of $\Hpj$ is quickly altered as a strong ejection of $\Hpj$ (negative $\dHpjdtown$) develops. After $t\sim 840$ $\dHpjdtown$ overcomes $\dHjdttrans$ and $\dHpjdt$ becomes negative: both $\Hj$ and $\Hpj$ decrease. As noted in \citet{Linan18}, $\Hj$ is not directly ejected but is first converted in $\Hpj$ and the later is ejected out of the simulation domain. This is the inverse process of what occurred during the driving phase, however occurring faster and more impulsively.

\subsection{Helicity eruptivity index} \label{sec:Hindex}

The evolution of the helicity eruptivity index, $\eta_H=\Hj/\Hv$ (\eq{etah}), is displayed in the middle panel of \fig{RelativeFractions} its values at a few selected times are given in \tab{EHvalues}.

For the Non-eruptive simulation, $\eta_H$ steadily increases during the driving phase until reaching $0.63$ at $t=300$ and then stays constant. For the Jet producing simulation, $\eta_H$ reaches $0.74$ at the end of its driving phase, at $t=360$. At the end of the driving phase, $\eta_H$ thus first presents a larger value (by 17\%) for the Jet producing simulation compared to the Non-eruptive one. This is to be compared with the free energy ratio, $\Ej/\E$, (see top panel of \fig{RelativeFractions} and \tab{EHvalues}) which maximum is also about 17\% higher for the Jet producing simulation relatively to the Non-eruptive simulation.

In the post-driving phase, while for the Non-eruptive simulation both $\eta_H$ and $\Ej/\E$ remain constant, the evolution of the helicity eruptivity index significantly differs from the free energy ratio for the Jet producing simulation. Once the driving has stopped, $\Ej/\E$ very slowly decreases. There is not significant evolution between the end of the forcing at $t=350$ and the generation of the jet after $t\sim 750$. On the contrary, $\eta_H$ further increases. Following the sensitive increase of $\Hj$ and the decrease of $\Hpj$ (cf. \sect{Hdyn}), $\eta_H$ goes from $0.74$ at $t=360$ to $0.8$ at $t=740$, before the generation of the jet. Said differently, at the onset of the jet $\Hj$ represents 80\% of the helicity content of the system. The helicity eruptivity index is at its peak value just before the onset of the eruptive behavior. During the generation of the jet, $\eta_H$ decreases and its value falls even below the value of the Non-eruptive simulation.

The increase of the helicity eruptivity index reveals the more and more dominating role that $\Hj$ has in $\Hv$. Another way to see this is to follow the ratio $\Hj/\Hpj$, as presented in \fig{RelativeFractions}. One note that for the Non-eruptive case, at the end of the driving phase $\Hj$ is about 1.7 times larger than $\Hpj$. In the case of the Jet producing simulation, $\Hj$ is about 3 times larger (2.88 at $t=360$) than $\Hpj$ at the end of the driving phase. This fraction further increases by 41\% during the post driving phase to reach $\Hj/\Hpj=4.07$ at $t=740$. The  $\Hj/\Hpj$ ratio presents a relative difference almost as important between the end of the driving phase and the onset of the jet, than the relative difference between the Non-eruptive case and the Jet producing case at the end of their respective driving phase. Again, the careful analyse of the helicity content reveals clearly the important dynamics/reconfiguration occurring in the system, which the magnetic energy is not able to capture.

It is interesting to see the role of $\Hj$ in conjunction with the eruptive behavior. \Fig{RelativeFractions} shows that at the onset of the generation of the jet, the relative helicity of the system is dominated by the non-potential helicity $\Hj$. This behavior is also observed in other numerical models. For example the Jet producing simulation of \citet{Pariat09a} analysed in \citet{Linan18} presents a similar decrease of $\Hpj$ as the system gets closer to the instability. Actually, in that simulation, $\Hpj$ even changes sign and has a chirality opposite to $\Hj$ and $\Hv$ \citep[cf. Figure 3 of][]{Linan18}. The helicity eruptivity index was thus larger than 1 at the onset of the jet. In the flux emergence simulations of \citet{Leake13b,Leake14a} studied in \citet{Pariat17}, eruption were generated for the systems which had the lower amount of $\Hpj$. The eruption were triggered the earliest in the system which contained an amount of $\Hpj$ of opposite sign to $\Hj$. It is therefore puzzling to see in the present study, that not only $\eta_H$ was higher after the point-of-no-return, that is higher at the end of the driving phase of the Jet producing case vs. the Non-eruptive case, but that during the post-driving phase, $\eta_H$ was further increasing, meaning $\Hj$ was further dominating $\Hpj$ as the system was approaching the actual eruption/generation of the jet. This highlights again the fact that helicities, $\Hj$, $\Hpj$ and $\eta_H$, seems to be tightly linked with the eruptive dynamics of solar-like active magnetic systems. 

\section{Conclusions and discussion} \label{sec:Conclusion}
\subsection{Summary} \label{sec:summary}

The present study is focussed on understanding the possible link between magnetic helicity and the eruptivity of solar-like magnetic systems. Here, we analyse the magnetic energy and helicity dynamics in two parametric 3D MHD numerical simulations that can induce solar coronal jets (\cf \sect{Sim}). In both simulations, the initial magnetic system is composed of a single 3D null point topology, dividing the domain in a closed field region (below the fan dome of the null point) and an open field region. In both cases, the system is driven by line-tied boundary motions inside the close domain, in order to form a flux rope initially contained within the closed domain (\cf \sect{Simforcing}). 

In one simulation (\cf \sect{SimJet}), presented and analysed in \citet{Wyper18b}, the system is driven sufficiently that a point-of-no-return is reached for the stability of the system : a jet is eventually generated following previous simulations results \citep{Pariat09a,Pariat15a,Pariat16,Wyper17}. Interestingly, the onset of the jet does not occur during or immediately after the driving phase in this simulation. The onset phase of the jet is significantly delayed after the end of the driving phase. During this post-driving phase, \citep[labelled "breakout phase" in][]{Wyper18b}, the flux rope presents a steady evolution involving reconnection which while sustained, is not eruptive/exponentially growing. During this phase, a reconfiguration of the magnetic field takes place: a fraction of the closed field magnetic flux becomes part of the flux rope, increasing its flux, while simultaneously the strapping field linking with it is removed. In a new simulation (\cf \sect{SimNonrupt}), the system is driven during a shorter time compared to the previously analysed one. During this shorter driving period, the point-of-no-return for the generation of the jet is not reached. The post-driving phase keeps on and is not followed by the onset of a jet. While some reconnection is present during the post driving phase, the flux rope created during the driving phase remains stable.

Because of their distinct behavior, it is particularly interesting to analyse the properties of the different magnetic energies and helicities (\cf \sect{Method}). We looked more specifically at the dynamics of the non-potential magnetic helicity, $\Hj$ (\eq{hpj}), of the volume threading magnetic helicity, $\Hpj$ (\eq{hpj}) and of the helicity eruptivity index $\eta_H$ (\eq{etah}). The later has been found in a few recent numerical models as well as in observations of solar active region to mark the eruptivity of the system \citet{Pariat17,Zuccarello18,Linan18,Moraitis19b,Thalmann21,Gupta21,Green22}.

The main results of our analyses are (\cf \sect{Results}):
\begin{itemize}
\item The driving motions during the driving phase induces the injection of free magnetic energy, $\Hj$, $\Hpj$, and hence the increase of both total magnetic energy and helicity. Since the driving phase lasts longer, more free energy, $\Hj$, total magnetic energy and helicity are injected in the Jet producing simulation. The Jet producing simulation is thus associated with a larger amount of $\E$, $\Efree$, $\Hj$ and $\Hpj$ compared to the Non-eruptive one, as expected from classical solar eruption theory. 
\item However, the volume threading helicity, $\Hpj$, is smaller at the end of the driving phase of the Jet producing case compared to the end of the driving phase of the Non-eruptive case. The additional forcing, during which the point-of-no-return is crossed, is coincident with this decrease of $\Hpj$.  
\item  During the post driving phases, $\Efree$ and $\E$ very slightly decrease in both simulations. Unlike magnetic energies, magnetic helicities are sensitive to the reconfiguration occurring in the post-driving phase of the Jet producing simulation. The fluxes of $\Hj$ and $\Hpj$ present completely distinct behaviors in each simulation during the post driving phase (\cf \sect{Hflux}). The helicities are thus able to uniquely capture key dynamics of the magnetic system to which the magnetic energies are blind.
\item During the post driving phases of the Jet producing simulation $\Hj$ and $\eta_H$ further increases. The onset of the jet is thus associated with a peak value of $\Hj$, $\eta_H$ and $\Hj/\Hpj$ (\cf \sect{Hindex}). These quantities are sensitively higher at the dawn of the onset of the jet compared to the end of the driving phase.
\end{itemize}

\subsection{Discussion} \label{sec:disc}

The first main outcome of this study relates to the comparative properties of magnetic helicity versus magnetic energy. As was already discussed in \citet{Linan20}, magnetic helicities appear to be significantly more sensitive quantities to the pre-eruptive properties of the magnetic system than magnetic energies. It is remarkable that magnetic energies are completely blind to the restructuring during the post-driving phase of the Jet producing simulation, while magnetic helicities do capture this evolution. It will also be worth investigating how the different restructuring dynamics (e.g. reconnection at the null point, reconnection below the flux rope, rise of the flux rope, ...) relates with the different observed changes in helicities. 

\begin{figure*}[ht!]
 \centering
\includegraphics[width=0.8\linewidth]{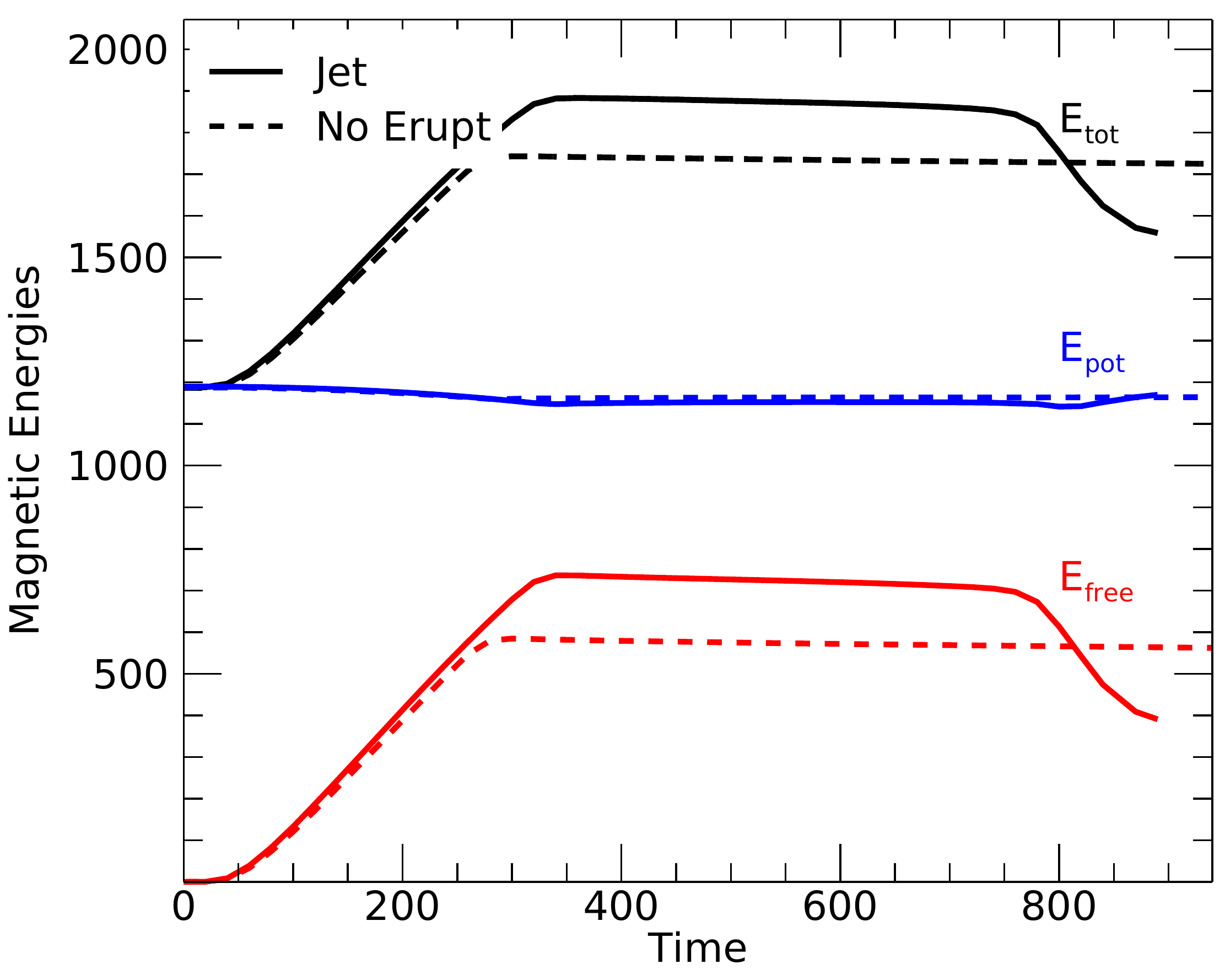}
\includegraphics[width=0.8\linewidth]{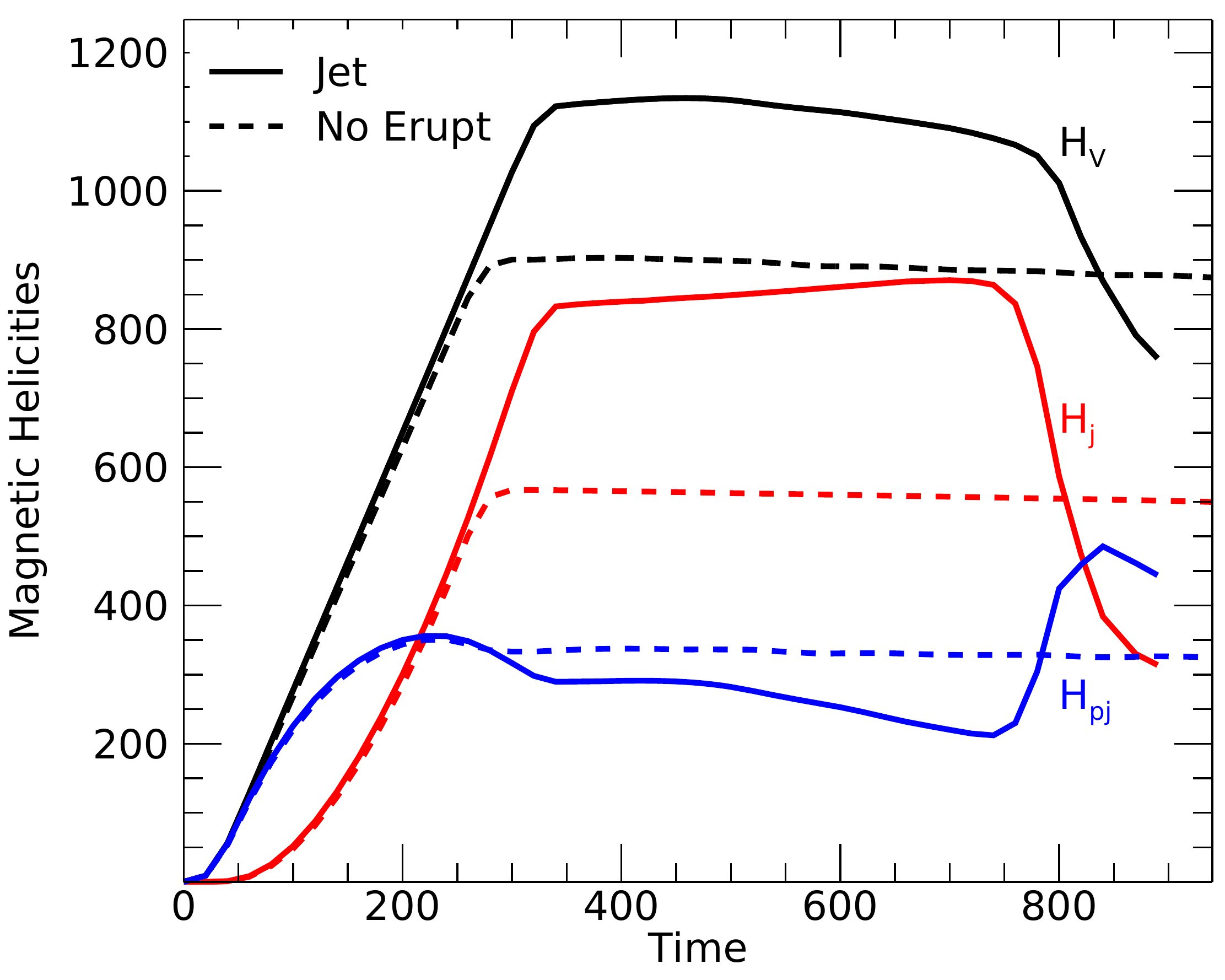}
 \caption{Top panel: evolution of the total magnetic energy ($\E$, black lines), potential magnetic energy ($\Ep$, blue lines), and free magnetic energy ($\Efree$, red lines) in the Non-eruptive (dashed lines) and in the Jet producing (continuous lines) simulations. Bottom panel: evolution of the total relative magnetic helicity ($\Hv$, black lines), non-potential magnetic helicity ($\Hj$, red lines) and volume-threading magnetic  helicity ($\Hpj$, blue lines),  in the Non-eruptive (dashed lines) and in the Jet producing (continuous lines) simulations.}
 \label{fig:EnergyHelicity}
 \end{figure*}

\begin{figure*}[ht!]
 \centering
\includegraphics[width=0.49\linewidth]{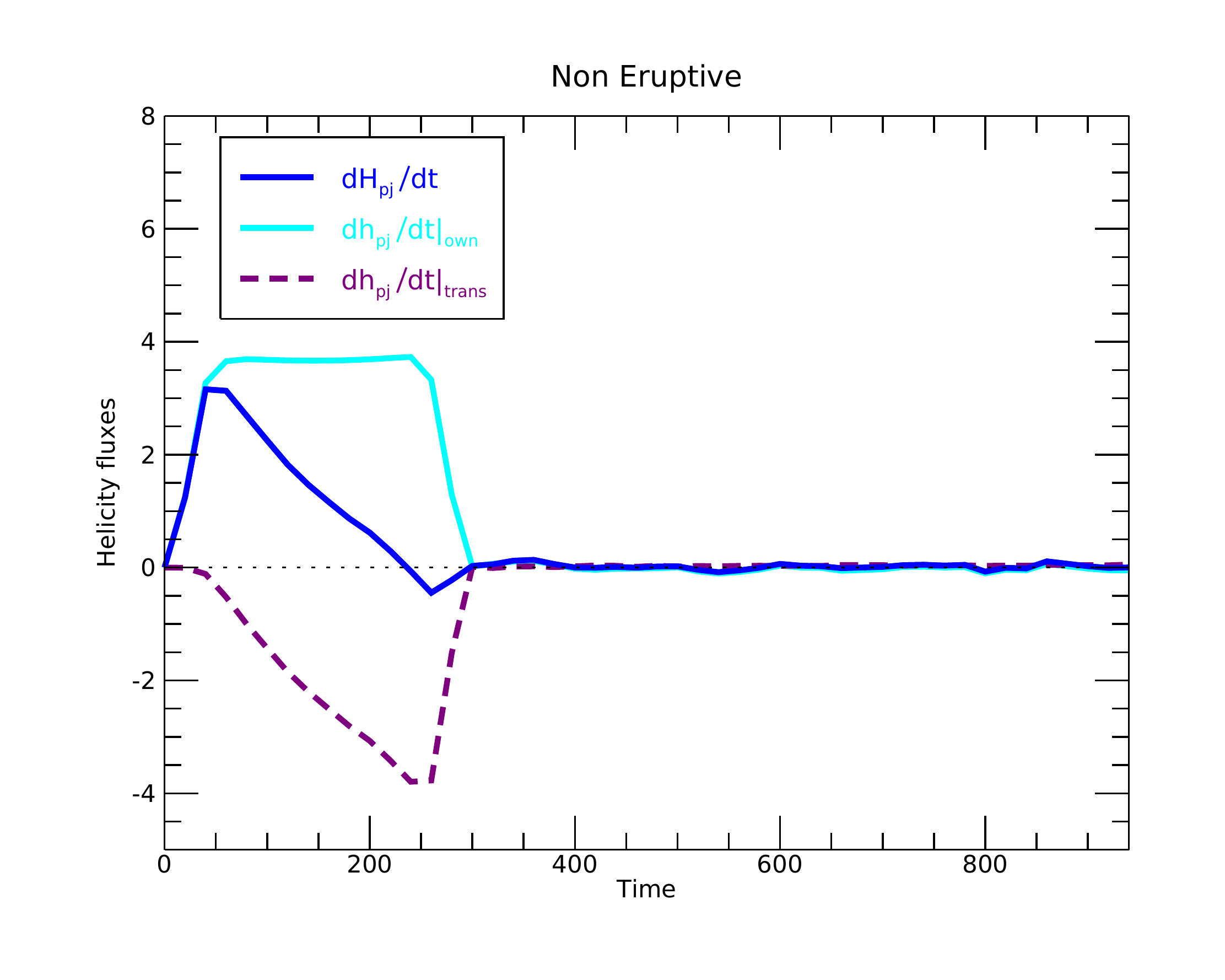}
\includegraphics[width=0.49\linewidth]{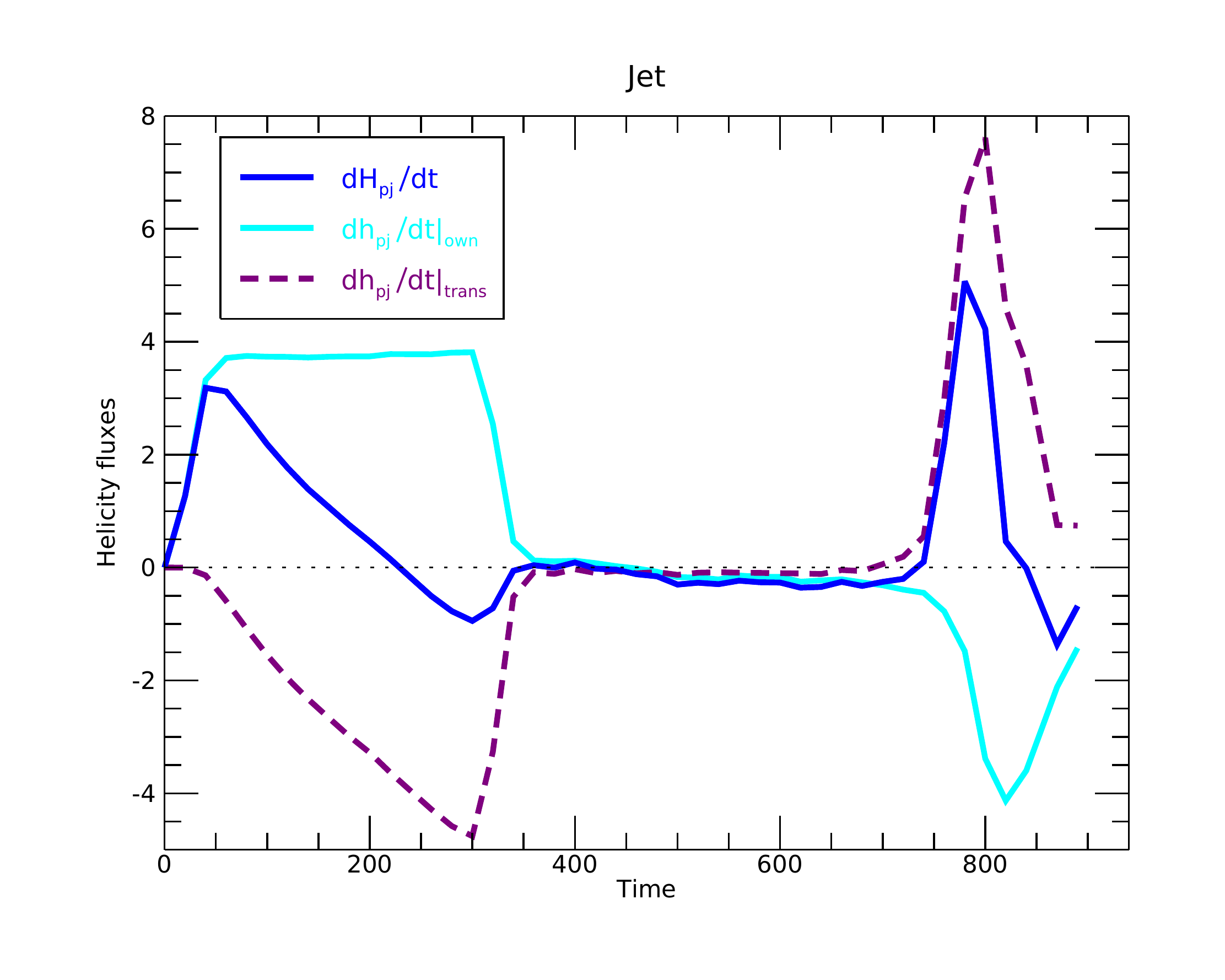}
\includegraphics[width=0.49\linewidth]{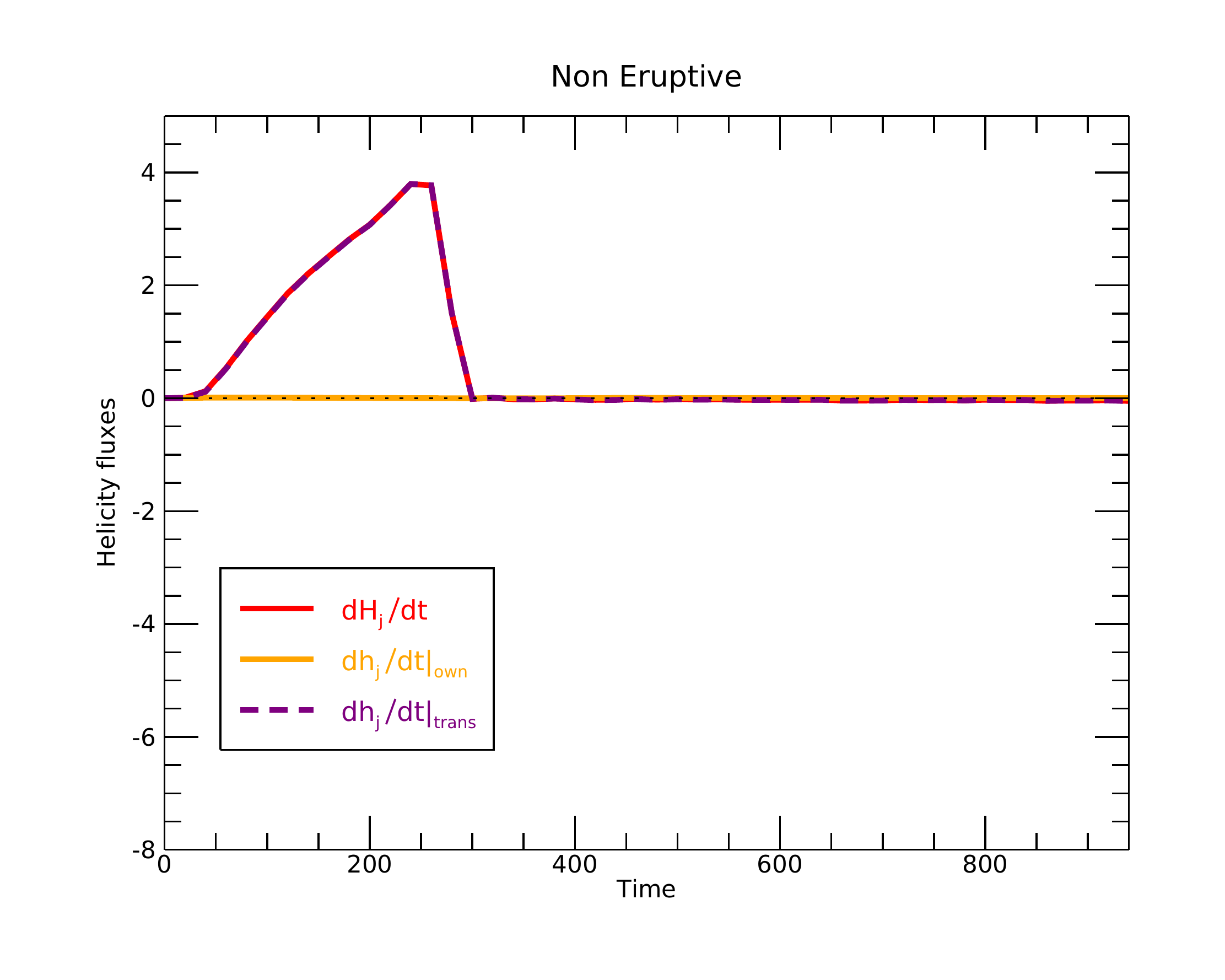}
\includegraphics[width=0.49\linewidth]{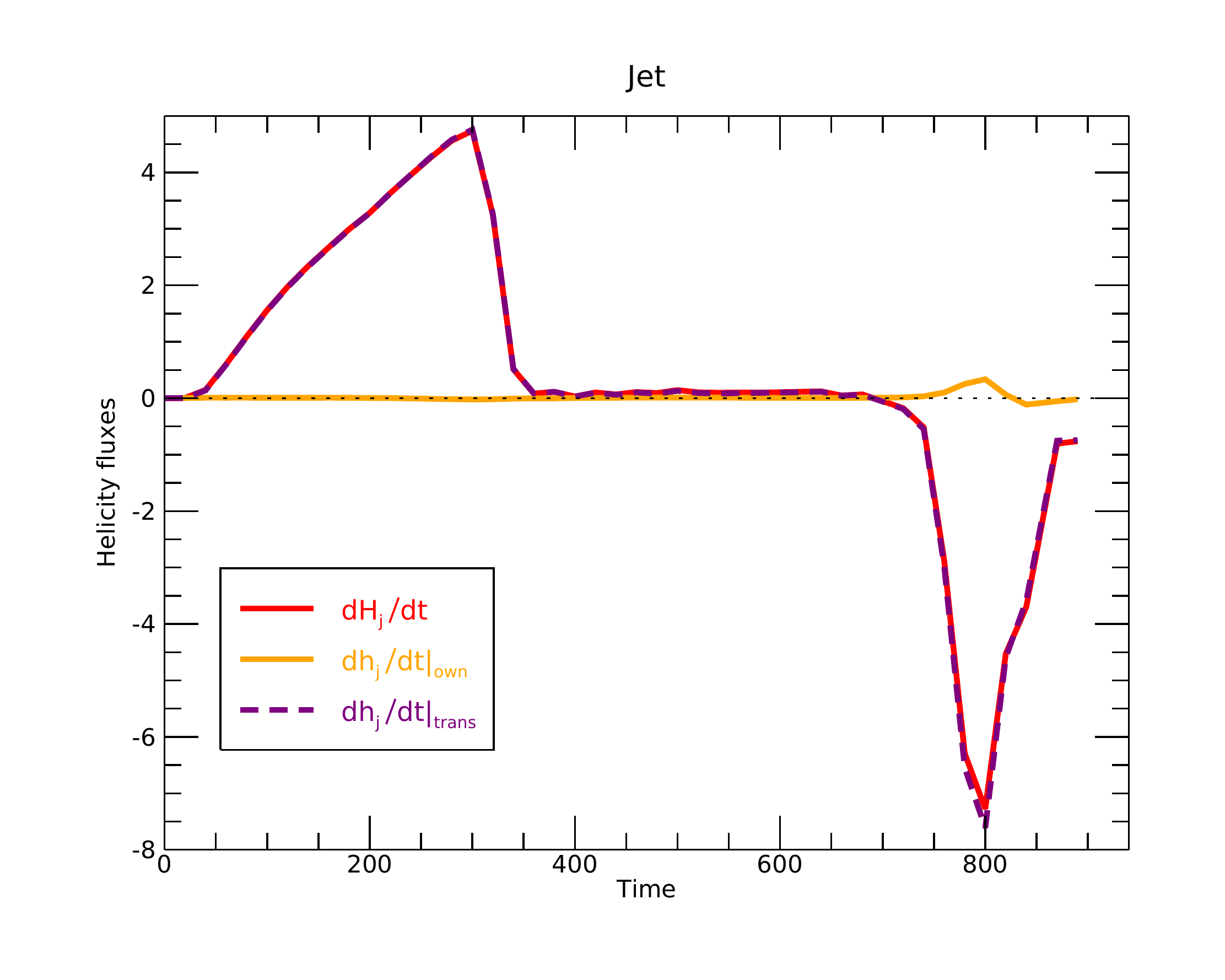}
\caption{Evolution of the terms of the time variation equation of $\Hpj$ (\eq{dhpjdt}, top panels) and $\Hj$ (\eq{dhjdt}, bottom panels) for the Non-eruptive (left column) and the Jet producing (right column) simulations: $\dHpjdt$ (blue line), $\dHpjdtown$ (cyan line), $\dHjdt$ (red line), $\dHjdtown$ (orange line), and $\dHpjdttrans=-\dHpjdttrans$ (purple dashed lines).}
 \label{fig:FluxesHjHpj}
 \end{figure*}

\begin{figure*}[ht!]
 \centering
\includegraphics[width=0.49\linewidth]{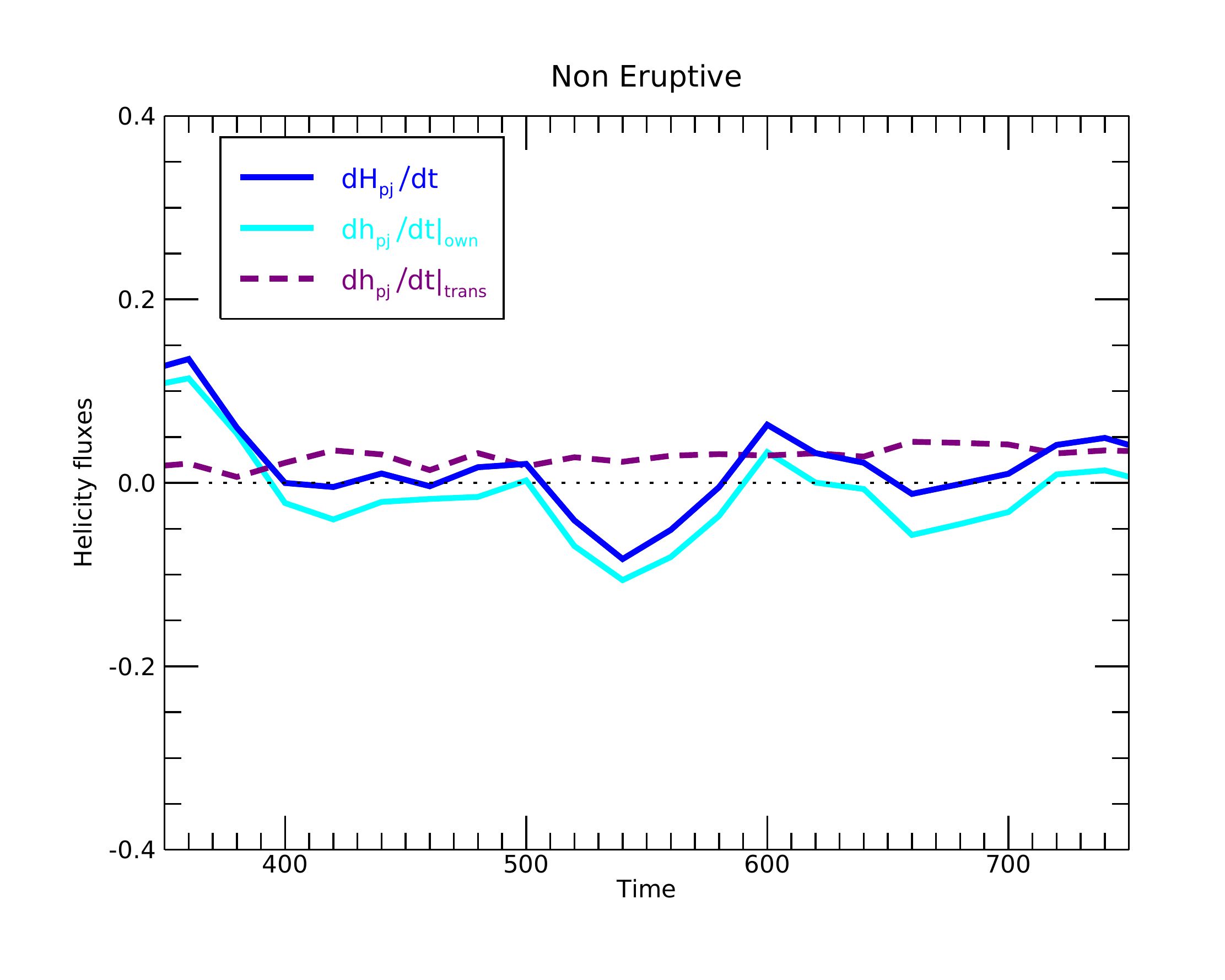}
\includegraphics[width=0.49\linewidth]{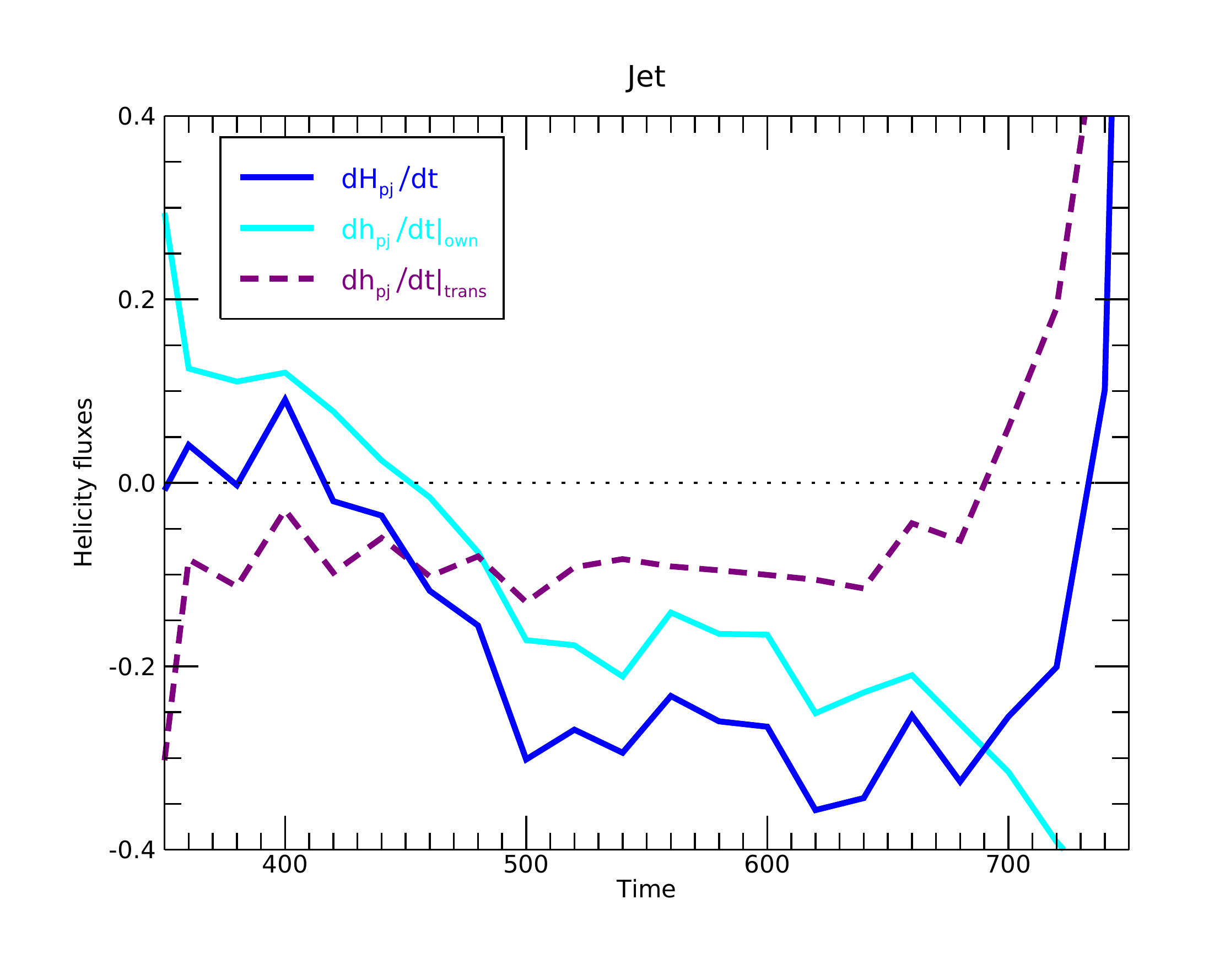}
\includegraphics[width=0.49\linewidth]{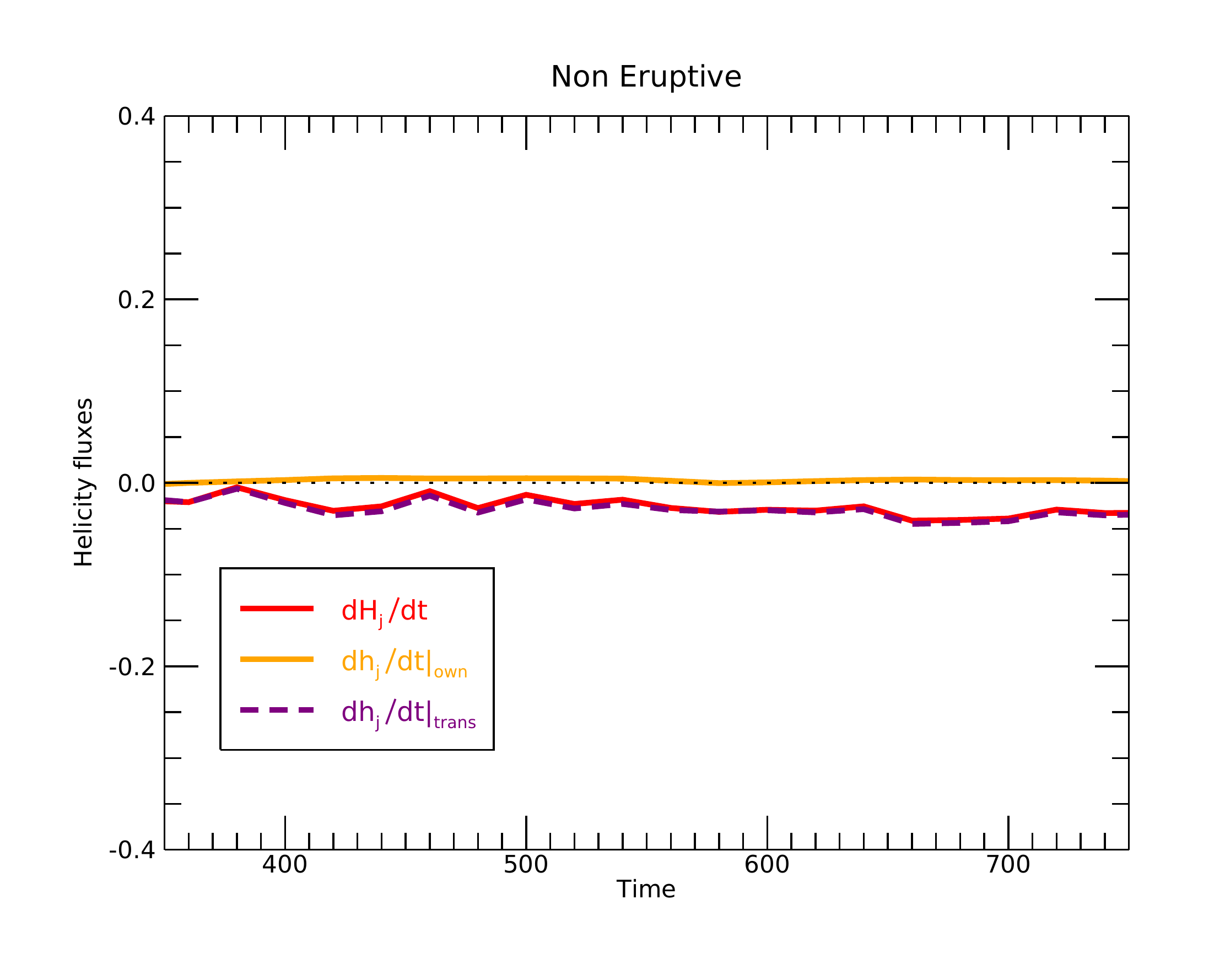}
\includegraphics[width=0.49\linewidth]{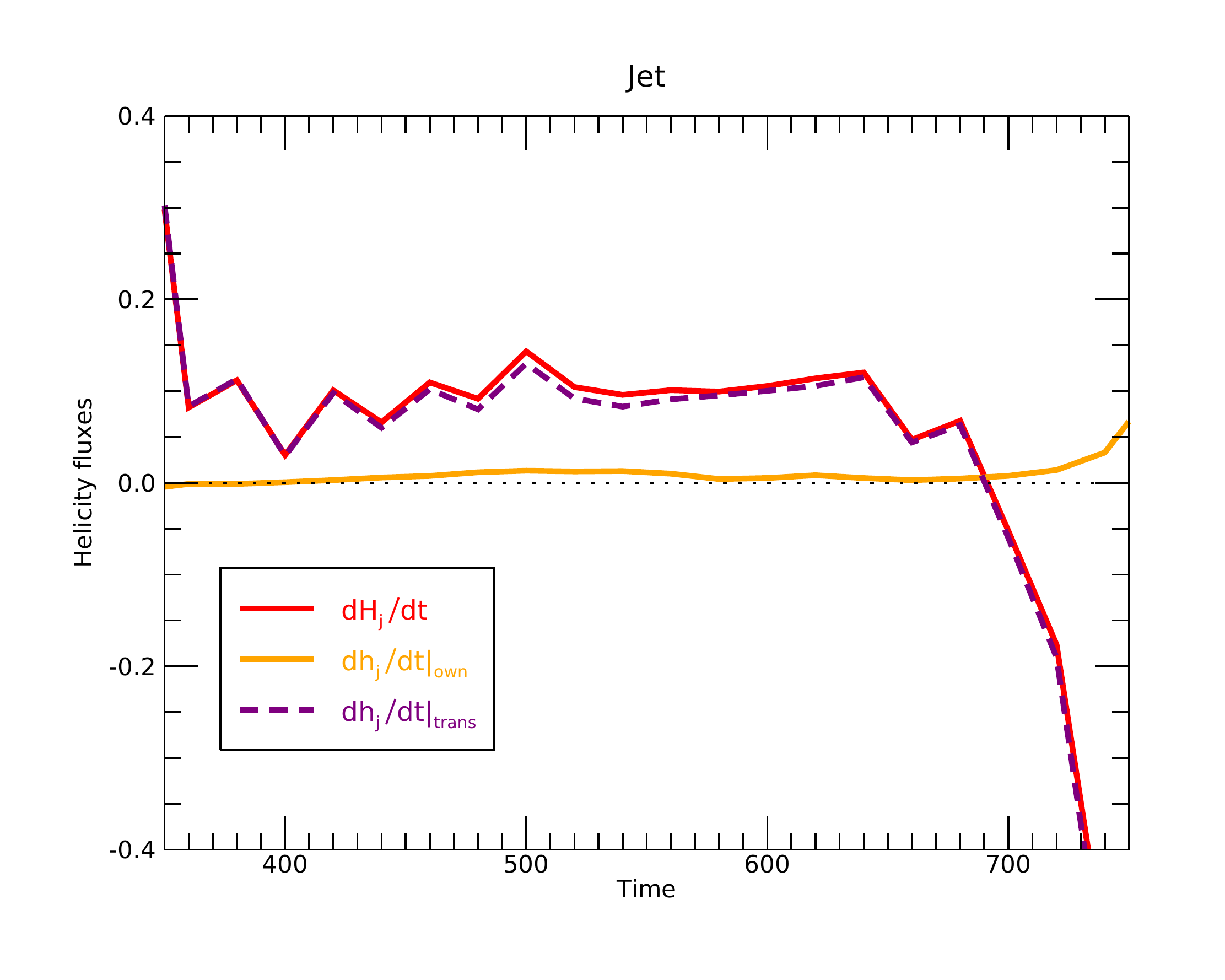}
 \caption{Same as \fig{FluxesHjHpj} but focussed on the post driving phase, between $t=350$ and $t=750$.}
 \label{fig:FluxesHjHpjzoom}
 \end{figure*}

\begin{figure*}
 \centering
\includegraphics[width=0.5\linewidth]{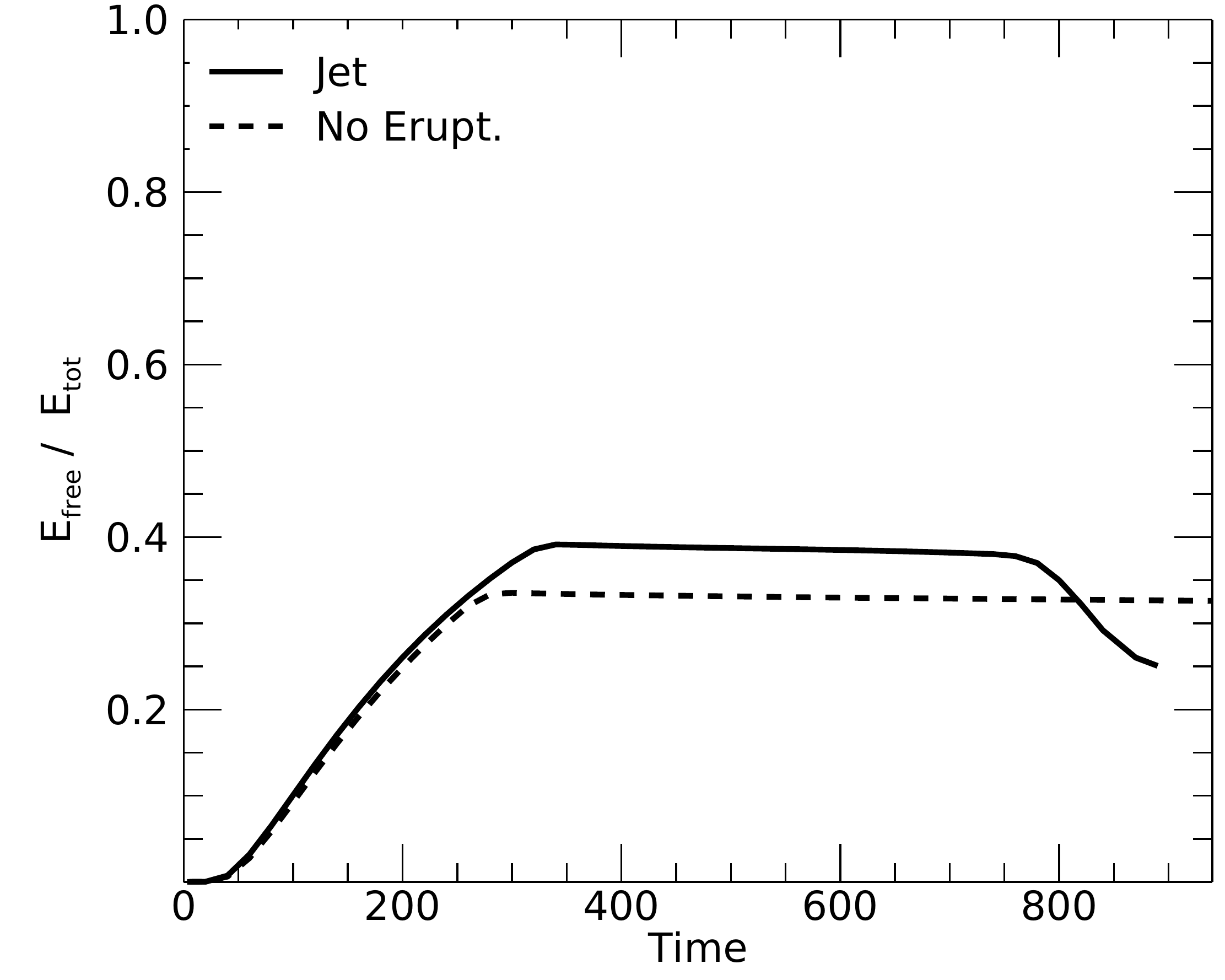}
\includegraphics[width=0.5\linewidth]{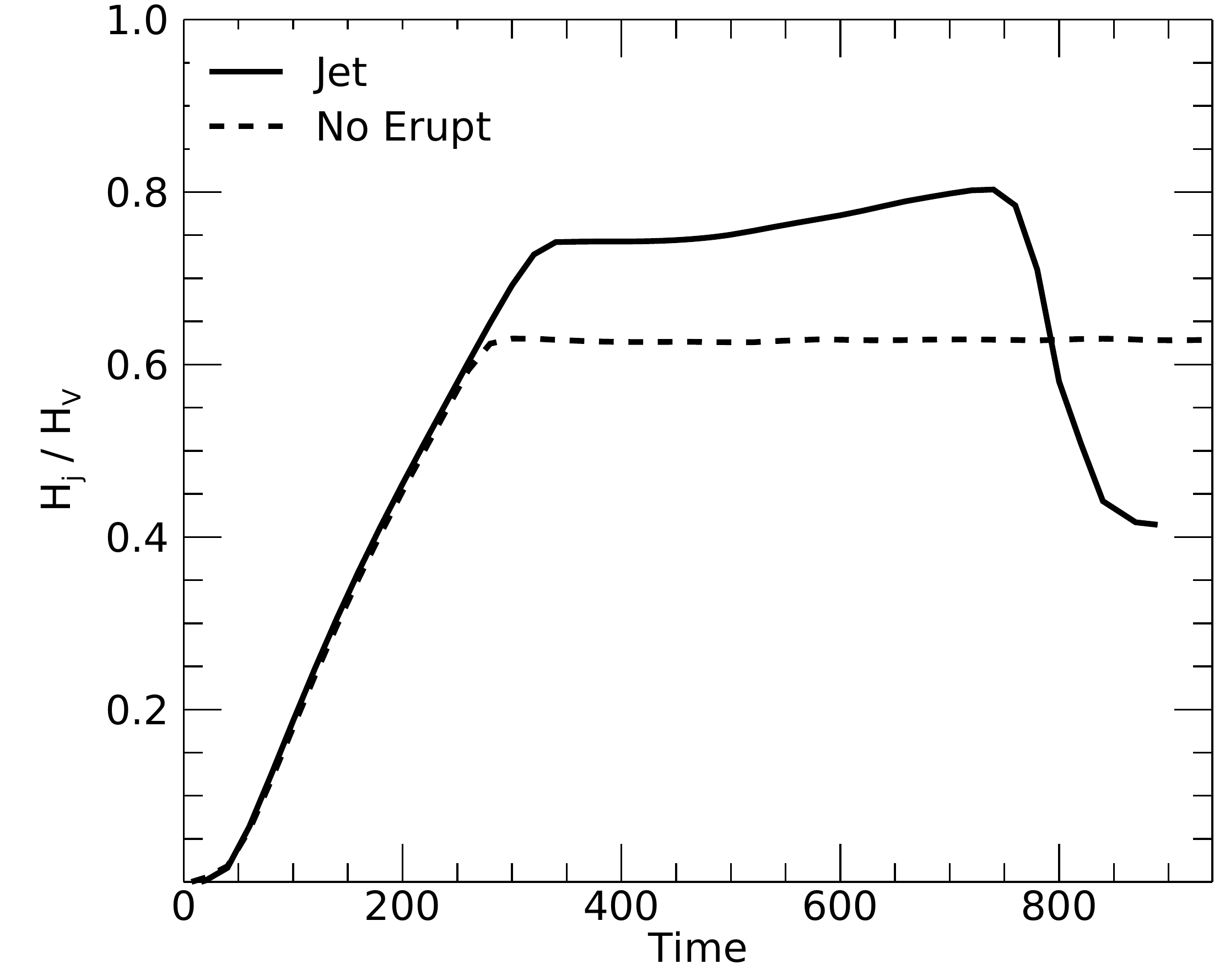}
\includegraphics[width=0.5\linewidth]{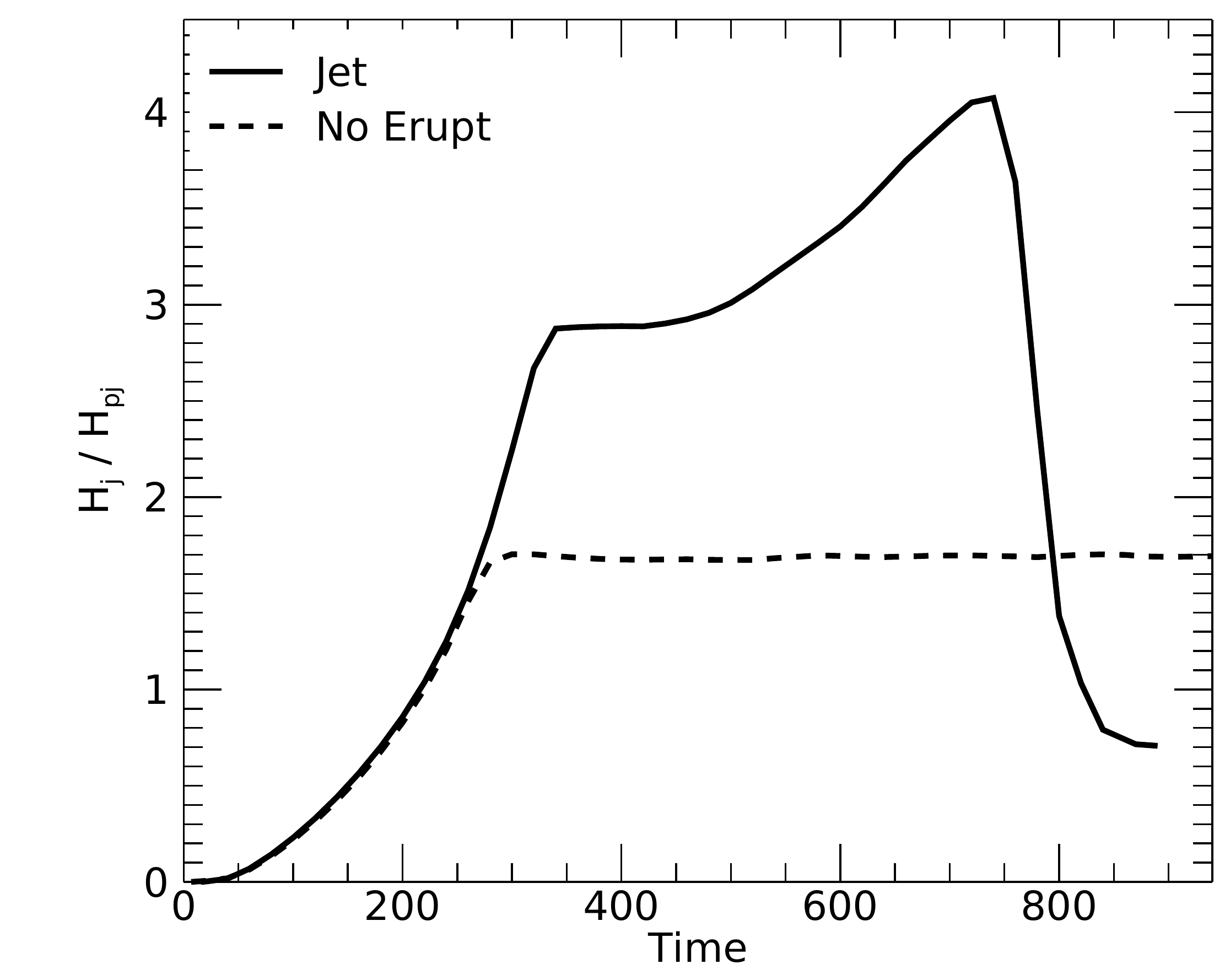}
 \caption{Time evolution of non-dimensional quantities in the Non-eruptive (dashed lines) and Jet producing (continuous lines) simulations: $\Efree/\E$ (top panel), helicity eruptivity index $\eta_H=\Hj/\Hv$ (middle panel), and $\Hj/\Hpj$ (bottom panel).
}
\label{fig:RelativeFractions}
\end{figure*}

Another feature worth mentioning, which likely requires further studies, is the fact that magnetic helicities seems to change earlier that magnetic energies before the onset of the eruptive dynamics. Indeed, one observes that $\Hj$ (resp. $\Hpj$) presents a maximum (resp. local minimum) at $t=740$ (see \fig{EnergyHelicity}). The decrease of the total and free magnetic energy related to the eruptive behavior only becomes observable after $t>760$. More strikingly, when looking at the fluxes (see \fig{FluxesHjHpjzoom}) of $\Hj$ and $\Hpj$, one observe that the transfer of helicity between $\Hj$ and $\Hpj$ reverts as early at $t=700$.  This obvious change of the helicity dynamics is likely related to the onset of generation of the jet. Helicities, and their fluxes, may thus constitute warning of the imminent onset of eruptive events.  

The second major outcome of the present analyse relates to the potential use of the helicity eruptivity index, $\eta_H$, in eruption prediction. The numerical experiments of \citet{Zuccarello18} showed clearly that the onset of the eruptive behavior was associated with a threshold in $\eta_H$. Unlike in \citet{Zuccarello18}, where the point-of-no-return was precisely determined, the present parametric simulations do not permit to completely link the moment in which the system becomes unstable with the helicities. Although the behaviour during the post-driving phase heavily involves breakout reconnection above the flux rope structure \citep{Wyper17,Wyper18b}, the present simulations do not permit the precise determination of which instability is triggering this eruptive behaviour, \ie whether it is a resistive instability, as argued by the "breakout" scenario \citep{Antiochos99} or an ideal MHD instability such as the Torus instability \citep{Kliem06,Aulanier10} that acts to kick off or to later supplement the eruptive evolution. Precisely determining this would require further parametric MHD simulations, perhaps alongside the use of an ideal code \citep[\eg][]{Rachmeler10}, which is beyond the scope of this investigation. What can be strictly said is that a point-of-no-return is crossed during the extra driving time of the Jet producing simulation, between $t=300$ and $t=350$, which eventually lead to the eruptive behavior.

 The observed delay between the point-of-no-return and the actual onset of the jet is however of high interest. Two scenarios can be hypothesised, which present numerical experiments cannot discriminate. In the first scenario, the trigger of the eruptive behavior occurs during the supplementary driving time of the Jet producing simulation. The post-driving phase can thus be viewed as a "linear" phase of the loss of equilibrium that inevitably leads to the eruptive generation of the jet. The jet onset after $t\sim740$ is then simply the exponential phase of the development of the instability initiated between $t=300$ and $t=350$. In the present simulation the linear phase is particularly long, enabling its analyse in detail. Since  $\eta_H$ is higher for the Jet producing case than the Non-eruptive case in this time period, this scenario does not contradict that $\eta_H$ is related to the instability trigger.
 
 In a second alternative scenario, the point-of-no-return is not directly associated with the trigger of the instability at the origin of the eruptive behavior. The point-of-no-return may here be associated with a first instability that induces the reconfiguration of the magnetic closed system with the further build-up of a flux rope. Doing so, helicity is further converted from $\Hpj$ to $\Hj$ in the Jet producing case, in opposition to the Non-eruptive case. As the magnetic system reconfigures itself, and  $\eta_H$ further rises, the system may be driving towards a second instability (ideal or not), this one directly associated with the onset of the eruption/jet. If $\eta_H$ is indeed associated with such eruptive instability, this would explain why the system erupts only after $t>740$ and not directly at the end of the driving phase. The threshold in $\eta_H$ may not yet have been reached at $t=350$ and it is only thanks to the reconfiguration in the post-driving phase, that $\eta_H$ reaches the instability threshold level.
 
Whichever scenario is correct, this study further confirms the results of \citet{Pariat17,Zuccarello18,Linan18,Linan20} pointing towards a tight link between the eruptivity of magnetic configurations and magnetic helicities, and in particular the helicity eruptivity index $\eta_H$. Similarly to the previously analysed simulations, we find in the present simulation that $\eta_H$ is higher for the Jet producing simulation compare to the Non-eruptive case, that $\eta_H$ present a peak just before the onset of the eruptive jet and that the value of $\eta_H$ has decrease once the eruption/jet occurred. 
  
However, the nature of the causal link between $\eta_H$ and the trigger of eruptions still needs to be determined. \citet{Pariat17,Zuccarello18} suggested than $\eta_H$ is related to the torus instability. Recently \citet{Kliem22} showed that kink and torus unstable system where very efficient at shedding magnetic helicity, and in particular $\Hj$, while $\Hpj$ was only partly extracted. They found that the systems were stable when $\eta_H$ lied below a certain threshold. This study highlights the possible link between $\eta_H$ and the torus instability.
 
 The search for the causal link between the properties of the pre-eruptive magnetic field and the trigger of active solar events is an extremely dynamic topic in solar physics \citep[\eg][]{Leka19a,Leka19b,ParkSH20,Georgoulis21}. Innovative quantities permitting an advance prediction of eruptive events are being looked for. Magnetic twist, winding and helicity, which all relates to the level of entanglement/complexity of the magnetic field  seems to constitute a promising approach. In addition the helicity eruptivity index on which this study focusses, other helicity related quantities have very recently been proposed. Historically, multiples studies have focussed on the total helicity content \citep[\eg][]{Nindos04,Labonte07,ParkSH10,Tziotziou12,Vemareddy19,Liokati22}. Recently, in a 2D parametric numerical study, \citet{Rice22} found that major eruptions were best predicted by thresholds in the ratios of rope current to magnetic energy of helicity. They noted that the helicity eruptivity index was negatively correlated with eruptions.  \citet{LiT22} has proposed to use the ratio of a twist parameter to the total unsigned flux to distinguish large eruptive and confine flares. Building on the theoretical studies of \citet{Prior20} and \citet{MacTaggart21}, \cite{Raphaldini22} have shown that magnetic winding could successfully indicate the flaring/eruptive activity in some active regions. All these results points, to the importance of twist/helicity in the physics of solar eruption. Because of the inherent difficulties to measure these quantities, and of the tricky properties of some \citep[\eg the non simple additivity of relative magnetic helicity][]{Valori20}, a vast effort must still be carried to identify truly meaningful quantities for flare and eruption prediction.

\begin{acknowledgements}
We thank the anonymous referee for her/his thorough review of the manuscript. This project is a direct outcome of the visit of EP and TP at Durham University supported by the Alliance Hubert Curien Program funded by the french Ministère de l'Europe et des Affaires Etrangères and the Ministère de l'Enseignement Supérieur et de la Recherche, and managed by Campus France. This work was supported by the French Programme National PNST of CNRS/INSU co-funded by CNES and CEA. EP also acknowledges financial support from the French national space agency (CNES) through the APR program. LL acknowledges support from the European Union’s Horizon 2020 research and innovation programme under grant agreement N$^o$ 870405. The numerical simulations were sponsored by allocations on Discover at NASA's Center for Climate Simulation and on the DiRAC Data Analytic system at the University of Cambridge, operated by the University of Cambridge High Performance Computing Service on behalf of the STFC DiRAC HPC Facility (www.dirac.ac.uk) and funded by BIS National E-infrastructure capital grant (ST/K001590/1), STFC capital grants ST/H008861/1 and ST/H00887X/1, and STFC DiRAC Operations grant ST/K00333X/1. DiRAC is part of the National E-Infrastructure.
\end{acknowledgements}

%
%


\bibliographystyle{aa} 
\bibliography{HeliPWJet.bib}



\begin{table*}
  \begin{tabular}{ccccccccccc}
Simulation & Time & $\E$ & $\Ep$ & $\Efree$ & $\Efree/\E$ & $\Hv$ & $\Hj$ & $\Hpj$ & $\eta_H$ & $\Hj/\Hpj$ \\
    \hline
 \multirow{7}{*}{Non-eruptive}  &    0  &       1187 &       1187 &     0.0 &   0.0 &    0.0 &  0.0 &    0.0 &   0.0 &   0.0 \\
      & 300  &       1743 &       1160 &       584 &      0.34 &       900 &       567 &       333 &      0.63 &       1.7 \\
      & 360  &       1741 &       1162 &       581 &      0.33 &       902 &       566 &       336 &      0.63 &       1.68 \\
      & 460  &       1738 &       1163 &       577 &      0.33 &       900 &       564 &       336 &      0.63 &       1.68 \\
      & 700  &       1731 &       1164 &       569 &      0.33 &       886 &       557 &       329 &      0.63 &       1.70 \\
      & 740  &       1730 &       1164 &       568 &      0.33 &       885 &       556 &       328 &      0.63 &       1.69 \\
      & 890  &       1726 &       1164 &       564 &      0.33 &       878 &       552 &       326 &      0.63 &       1.69 \\
    \hline
     \multirow{7}{*}{Jet producing} & 0  &       1187 &       1187 & 0.0 & 0.0 &    0.0 &  0.0 &  0.0 &   0.0 &  0.0 \\
     & 300 &       1832 &       1155 &       678 &    0.37 &       1027 &       710 &       317 &      0.69 &       2.24 \\
     & 360  &       1883 &       1149 &       736 &      0.39 &       1126 &       836 &       290 &      0.74 &       2.88 \\
      & 460 &       1879 &       1152 &       729 &      0.39 &       1134 &       845 &       289 &      0.75 &       2.92 \\
      & 700  &       1861 &       1152 &       711 &      0.38 &       1091 &       871 &       220 &      0.80 &       3.96 \\
      & 740  &       1853 &       1151 &       705 &      0.38 &       1076 &       864 &       212 &      0.80 &       4.07 \\
      & 890  &       1558 &       1170 &       391 &      0.25 &       758 &       314 &       444 &      0.41 &      0.71 \\
    \hline
  \end{tabular}
\caption{Values of magnetic energies and helicities, and some of their ratio at different instant of the simulations.}
\label{tab:EHvalues}
\end{table*}


%

 




\end{document}